\documentclass[prx,superscriptaddress,floatfix,showpacs,notitlepage,twocolumn]{revtex4-2}
\usepackage[latin9]{inputenc}
\usepackage{color}
\usepackage{textcomp}
\usepackage{amsmath}
\usepackage{amssymb}
\usepackage{mathtools}
\usepackage{graphicx}
\usepackage{esint}
\usepackage{framed} 
\usepackage[dvipsnames]{xcolor}
\usepackage{hyperref}
\usepackage{enumerate}
\usepackage{enumitem}

\usepackage{moreenum}
\hypersetup{
  linkcolor = black,
  citecolor  = blue,
  urlcolor = black,
  colorlinks = true,
}
\makeatletter

\usepackage{amsfonts}
\usepackage{bbold}
\usepackage{natbib}
\usepackage{svg}
\usepackage{float}
\usepackage[export]{adjustbox}
\hypersetup{    colorlinks,    linkcolor={blue!50!black},    citecolor={blue!50!black},    urlcolor={blue!80!black}}
\usepackage[all]{hypcap}
\usepackage{bbold}
\graphicspath{{img/}}

\newcommand*{\bra}[1]{\ensuremath{\langle #1 \vert}}
\newcommand*{\ket}[1]{\ensuremath{\vert #1 \rangle}}
\newcommand{\braket}[2]{\langle #1 | #2 \rangle}

\newcommand{\lr}[1]{\left( #1 \right)}

\renewcommand{\vec}[1]{{\boldsymbol{#1}}}

\begin{document}

\title{{Hamilton-Jacobi-Bellman equations for Rydberg-blockade processes}}

\author{Charles Fromonteil}\email{Charles.Fromonteil@uibk.ac.at}
\affiliation{Institute for Theoretical Physics, University of Innsbruck, 6020 Innsbruck, Austria}
\affiliation{Institute for Quantum Optics and Quantum Information of the Austrian Academy of Sciences, 6020 Innsbruck, Austria}
\author{Roberto Tricarico}\email{r.tricarico@ssmeridionale.it}
\affiliation{Institute for Theoretical Physics, University of Innsbruck, 6020 Innsbruck, Austria}
\affiliation{Institute for Quantum Optics and Quantum Information of the Austrian Academy of Sciences, 6020 Innsbruck, Austria}
\affiliation{Scuola Superiore Meridionale, Largo San Marcellino 10, 80138 Napoli, Italy}
\author{Francesco Cesa}
\affiliation{Institute for Theoretical Physics, University of Innsbruck, 6020 Innsbruck, Austria}
\affiliation{Institute for Quantum Optics and Quantum Information of the Austrian Academy of Sciences, 6020 Innsbruck, Austria}
\affiliation{Department of Physics, University of Trieste, Strada Costiera 11, 34151 Trieste, Italy}
\affiliation{Istituto Nazionale di Fisica Nucleare, Trieste Section, Via Valerio 2, 34127 Trieste, Italy}
\author{Hannes Pichler}
\affiliation{Institute for Theoretical Physics, University of Innsbruck, 6020 Innsbruck, Austria}
\affiliation{Institute for Quantum Optics and Quantum Information of the Austrian Academy of Sciences, 6020 Innsbruck, Austria}

\begin{abstract} 
We discuss time-optimal control problems for two setups involving globally driven Rydberg atoms in the blockade limit by deriving the associated Hamilton-Jacobi-Bellman equations.
From these equations, we extract the globally optimal trajectories and the corresponding controls for several target processes of the atomic system, using a generalized method of characteristics. We apply this method to retrieve known results for CZ and C-phase gates, and to find new optimal pulses for all elementary processes involved in the universal quantum computation scheme introduced in [Physical Review Letters {\bf131}, 170601 (2023)].
\end{abstract}
\maketitle
\section{Introduction}
\vspace{-3mm}
Optically trapped neutral atoms are a promising platform for quantum computation due to their controllability and scalability~\cite{saffmanQuantumInformationRydberg2010,browaeysManybodyPhysicsIndividually2020,Henriet2020quantumcomputing,morgado2021RydbergQubits,kaufmanQuantumScienceOptical2021,weimerRydbergQuantumSimulator2010}. 
By means of optical tweezers, atoms can be arranged into arrays, and qubit degrees of freedom can be encoded in the electronic states of the atoms~\cite{norcia2024iterative,jenkins2022YtterbiumSpinqubits,singh2022DualElement}. These electronic states are manipulated by driving the system with laser fields, enabling quantum information processing~\cite{ebadiQuantumOptimizationMaximum2022a,byunFindingMaximumIndependent2022,ebadi2021QuantumPhases,scholl2021SimulationAntiferromagnets,steinertSpatiallyTunable,hollerith2022DistanceSelectiveInteractions,anand2024DualSpecies}. In particular, multiqubit operations can be realized by exploiting the strong van der Waals interactions among high principal quantum number states~\cite{levineParallelGates,madjarovHighfidelityEntanglementDetection2020,grahamMultiqubitEntanglementAlgorithms2022,bluvsteinQuantumProcessorBased2022,fu2022HighFidelityEntanglement,mcdonnell2022DemonstrationGateEIT,ma2023HighFidelityGates,everedHighFidelityParallel}, via the so-called Rydberg blockade mechanism, which prevents simultaneous excitation of nearby atoms~\cite{jakschRydbergGates,lukinDipoleBlockadeQuantum2001,urbanObservationRydbergBlockade2009,wilkEntanglementTwoIndividual2010}. This has led to intense research concerning the design and optimization of Rydberg-blockade-induced processes in recent years \cite{janduraTimeOptimalTwoThreeQubit2022,paganoErrorbudgetingControlledphaseGate2022,shi2018DeutschToffoliCNOT,Wu2022NatComm,fromonteilProtocolsRobustness,janduraOptimizingGates,mohanRobustControl,chang2023RydbergCZOptimal,saffman2020SymmetricRydbergCZ,pelegri2022GatesAdiabatic,mitra2023EntanglementAdiabaticRydberg}.

Improving multiqubit processes' fidelity is of crucial importance both for the implementation of deep quantum circuits in near-term devices~\cite{preskillQuantumNISQ} and to achieve fault tolerance~\cite{shor1996FaultTolerant,dennis2002Topological,fowlerSurfaceCodes,bluvstein2023Logical}. An important source of errors in neutral atom platforms is the finite lifetime of excited states~\cite{bluvsteinQuantumProcessorBased2022}, typically leading to error rates that grow with the process duration \cite{paganoErrorbudgetingControlledphaseGate2022}. This underlines the importance of designing time-optimal protocols. 
Indeed, recent efforts have achieved remarkable results optimizing pulse shapes for entangling gates, either through gradient descent methods based on time discretization or through ansatz-based optimization ~\cite{janduraTimeOptimalTwoThreeQubit2022,everedHighFidelityParallel}.
However, assessing the \textit{global} optimality of protocols obtained with these methods is challenging and requires a different approach.

In this work, we employ the Hamilton-Jacobi-Bellman (HJB) formalism as a framework for seeking global optima in time-optimal control problems with globally driven Rydberg arrays. 
The Bellman principle of optimality \cite{bellman1957dynamic} allows to cast continuous-time optimal control problems in the form of a nonlinear first-order partial differential equation (PDE), the HJB equation~\cite{kirkOptimalControlTheory}. This equation admits a unique generalized viscosity solution, which gives the global optimum of the problem \cite{crandallViscositySolutions,crandallPropertiesViscosity}, and from which optimal trajectories and controls can be directly generated. Solving such a PDE in high-dimensional state spaces is generally challenging due to the so-called ``curse of dimensionality". Here, we address this problem by identifying a minimal set of relevant state variables in globally driven Rydberg systems and derive a low-dimensional expression of the HJB equation, which can be integrated by means of a generalized method of characteristics \cite{subbotinaHJCharacs,subbotinMinimax,bayenMinimax}.

We apply this approach to time-optimal control problems in globally driven Rydberg arrays in two different physical settings. 
On the one hand, we optimize protocols to implement Rydberg entangling gates, confirming the global optimality of previous results on controlled-phase gates \cite{janduraTimeOptimalTwoThreeQubit2022,everedHighFidelityParallel}. 
For this, we focus on global driving protocols, since they are advantageous for scalability and fault-tolerant processor design \cite{bluvstein2023Logical}. 
On the other hand, we investigate processes involving the simultaneous control of multiple clusters of mutually blockaded atoms. Cluster dynamics play a central role in several recently proposed quantum simulation and computation protocols \cite{cesaUnivQCGlobal,maskara2023programmable}. In this work, we mainly focus on their use in the 
universal quantum computation scheme introduced in Ref.~\cite{cesaUnivQCGlobal}, and compute time-optimal protocols for all its fundamental building blocks. In this proposal, global driving is an integral feature of the processor design. Although the two settings correspond to different modes of operating atom arrays, and thus to different target processes, we show that they can be treated within a unified mathematical framework.

This work is organized as follows. In Sec.~\ref{sec:model}, we describe the two main settings (namely, the gate setting and the cluster one) to which the present analysis applies, and show how their dynamics can be reduced to that of independent effective two-level systems (TLSs). In Sec.~\ref{sec:HJB}, we apply the Bellman principle of optimality to the associated time-optimal control problems and derive the corresponding Hamilton-Jacobi-Bellman equations. In this section, we furthermore present a generalization of the method of characteristics to compute the solution of such equations. In Sec.~\ref{sec:2qubit}, we apply the HJB method to the canonical case of two effective TLSs, and optimize several quantum computation processes. Eventually, in Sec.~\ref{sec:3qubit}, we extend the analysis to three effective TLSs and similarly optimize relevant dynamical processes. 

\section{Model}\label{sec:model}
\begin{figure}
    \centering
    \includegraphics[width=\linewidth]{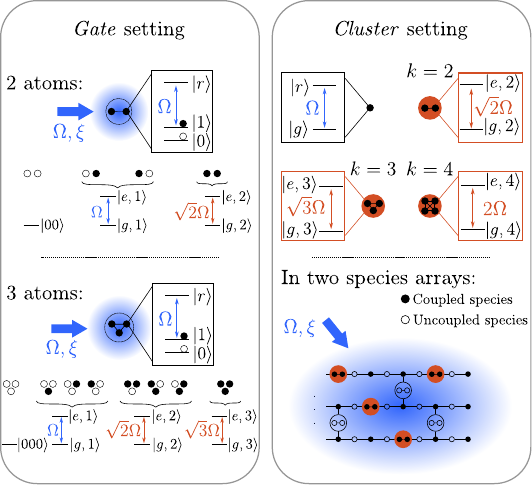}
    \caption{Graphical representation of the two physical settings of interest. In the \textit{gate} setting, atoms are placed within a blockade radius. The logical states of the qubits, $\ket{0}$ and $\ket{1}$, are encoded in two long-lived levels of the atoms, while an auxiliary Rydberg level $\ket{r}$ is exploited to perform gate operations. The atoms are collectively driven by a laser of Rabi frequency $\Omega$ and phase $\xi$, selectively driving the $\ket{1}\leftrightarrow\ket{r}$ transition.
    In the \textit{cluster} setting, two-level atoms, with ground state $\ket{g}$ and Rydberg state $\ket{r}$, are arranged in multiple isolated clusters (i.e., sets of atoms within a blockade radius). All clusters are simultaneously driven by a single laser of Rabi frequency $\Omega$ and phase $\xi$, tuned to the $\ket{g}\leftrightarrow\ket{r}$ transition. As shown in Ref.~\cite{cesaUnivQCGlobal}, universal quantum computation can be achieved in this setting by ``printing" a quantum circuit in the arrangement of a two-species atom array. In both settings, the dynamics decomposes into that of several independent effective two-level systems with enhanced Rabi frequency $\sqrt{n_k}\:\Omega$, where $n_k$ is the number of blockaded atoms that couple to the Rydberg state.}
    \label{fig:pres}
\end{figure}

\vspace{-3mm}
In this section, we introduce an effective model that formally describes Rydberg blockade dynamics in two distinct physical scenarios, namely (\textit{i}) in entangling gate 
design \cite{levineParallelGates}, and (\textit{ii}) in computation schemes based on independent atom clusters \cite{cesaUnivQCGlobal}. In the following, we refer to these two scenarios as \textit{gate} setting and \textit{cluster} setting, respectively (see Fig.~\ref{fig:pres}). 

As detailed below, both these scenarios can be reduced to a unified model consisting of $N$ independent and inequivalent two-level systems (TLSs). We index these effective TLSs by $k\in\{1,\dots,N\}$ and denote their ground and excited states by $\ket{g,k}$ and $\ket{e,k}$, respectively. Each of these TLSs responds to a time-dependent control field of phase $\xi(t)$ and Rabi frequency $\sqrt{n_k}\,\Omega(t)$, i.e., they are driven simultaneously but respond with different rates. That is, we consider $N$ independent Hamiltonians of the form:
\begin{equation}\label{eq:Hks}
    H_k(t) = \frac{\sqrt{n_k}\,\Omega(t)}{2}\left(e^{i\xi(t)}\ket{g,k}\bra{e,k}\;+\text{h.c.}\right).
\end{equation}
While the global parameters $\Omega(t)$ and $\xi(t)$ are controllable, the $n_k$ are fixed, defining individual but correlated effective Rabi frequencies. 
In this work, we aim at designing controls $(\Omega(t),\xi(t))$ that simultaneously realize given target processes on these TLSs. While this problem is already challenging by itself, we moreover seek its time-optimal solution, i.e., the fastest one.

In the remainder of this section, we elaborate on the physical models underlying the gate and cluster settings and discuss how their description reduces to Eq.~\eqref{eq:Hks}. 

\vspace{-3mm}
\subsection{The gate setting}\label{sec:gate}
\vspace{-3mm}
In this setting, we consider $N$ mutually blockaded three-level atoms. Each atom encodes a qubit in two long-lived, noninteracting levels (e.g. hyperfine states), $\ket{0}$ and $\ket{1}$; while an auxiliary Rydberg level $\ket{r}$ is utilized to mediate interactions for entangling operations. We are interested in realizing multiqubit gates by collectively driving the atoms with a single laser tuned to the $\ket{1}\leftrightarrow\ket{r}$ transition. This is a typical setup for implementing entangling gates with neutral atoms \cite{levineParallelGates}. Specifically, the dynamics is described by the following Hamiltonian:
\begin{equation} \label{eq:hamiltonian}
    H=\frac{\Omega(t)}{2}\sum_{j=1}^N\left(e^{i\xi(t)}\ket{1_j}\bra{r_j}+{\rm h.c.}\right)+\sum_{j<m}V_{jm}\hat{\mu}_j\hat{\mu}_m,
\end{equation}
with $j$ and $m$ indexing the atoms, $\hat{\mu}_j=\ket{r_j}\bra{r_j}$, and $V_{jm}$ denoting the strength of the interaction between atoms $j$ and $m$. Here, $\Omega(t)$ and $\xi(t)$ are the (positive) Rabi frequency and the phase of the driving laser, respectively. We note that this setting is sometimes formulated taking the laser detuning $\Delta$ (i.e., the derivative of $\xi$) as a control knob instead of the laser phase $\xi$ (e.g., in Ref.~\cite{levineParallelGates}). 
These two pictures are equivalent up to a change of reference frame, as discussed in Appendix~\ref{App:phasevsdetuning}.

In the so-called blockade limit ($V_{jm}\gg \Omega(t)$ for all pairs), multiple Rydberg excitations are dynamically forbidden. The Hamiltonian becomes block-diagonal, and the dynamics decomposes into that of independent effective two-level systems (TLSs). Indeed, apart from the $\ket{0}^{\otimes N}$ state, which is trivially invariant, each computational basis state forms a closed TLS with an excited state containing a Rydberg excitation delocalized over the coupled atoms. For instance, in the case of three atoms, the state $\ket{0,1,1}$ couples to ${(\ket{0,r,1}+\ket{0,1,r}})/\sqrt{2}$, while the state $\ket{0,1,0}$ couples to $\ket{0,r,0}$. The effective TLSs correspond to a partition of the Hilbert space into subspaces that do not mix during the dynamics.

More precisely, all computational basis states with the same Hamming weight $k$ (i.e., with $k$ atoms in state $\ket{1}$) undergo the same two-level dynamics. Therefore, we only distinguish between states with different $k$, and accordingly define $N$ inequivalent effective TLSs indexed by $k\in\{1,\dots,N\}$. Each of these effective TLSs consists of a state with no Rydberg excitation, ${\ket{g,k}=\ket{1}^{\otimes k}}$, and a state with a single one, ${\ket{e,k}=\tfrac{1}{\sqrt{k}}\sum_{l=1}^k\ket{1}^{\otimes (l-1)}\otimes\ket{r}\otimes\ket{1}^{\otimes (k-l)}}$ 
\footnote{
Here, we omit atoms in state $\ket{0}$, as they trivially do not evolve under the Hamiltonian in Eq.~\eqref{eq:hamiltonian}. The two states $\ket{g,k}$ and $\ket{e,k}$ are representative of the equivalent dynamics of all two-level subspaces corresponding to a computational basis state with Hamming weight $k$. For instance, the dynamics between $\ket{0,1}$ and $\ket{0,r}$ is the same as that between $\ket{1,0}$ and $\ket{r,0}$.}. 
We refer to these two states as the effective ground and excited states of the TLS, respectively. 
The TLSs do not couple to one another, and their dynamics is governed by Eq.~\eqref{eq:Hks}, with $n_k=k$. 
Expressing the state of the $k$-th TLS as ${\ket{\psi_k}=\psi_{g,k}\ket{g,k}+\psi_{e,k}\ket{e,k}}$, the Schr\"odinger equation returns:
\begin{align}
\begin{split}
    \dot{\psi}_{g,k}&=\frac{\sqrt{n_k}\,\Omega(t)}{2i}e^{i\xi(t)}\psi_{e,k},\\
    \dot{\psi}_{e,k}&=\frac{\sqrt{n_k}\,\Omega(t)}{2i}e^{-i\xi(t)}\psi_{g,k}\label{eq:Schrodinger}.
\end{split}
\end{align} 

The central goal in this setting is to engineer entangling gates. As unitary operations within the qubit subspace, those are uniquely defined by their action on the computational basis states, i.e., the effective ground states of the TLSs. Due to the independent evolution of the TLSs, the available unitaries are those that map each computational basis state to itself up to a phase. Thus, realizing a given gate corresponds to simultaneously driving the $N$ effective ground states accordingly. For instance, for the standard CZ gate, the $\ket{1,1}$ state acquires a $\pi$ phase, while the others remain unchanged: $\ket{z_1,z_2}\rightarrow(-1)^{z_1z_2}\ket{z_1,z_2}$, for $z_1,z_2\in\{0,1\}$. 
In the effective TLS formalism, this corresponds to the following process: $\ket{g,1}\rightarrow\ket{g,1}$, $\ket{g,2}\rightarrow-\ket{g,2}$ (see Sec.~\ref{Sec:results}).
\vspace{-3mm}
\subsection{The cluster setting}\label{sec:cluster}
\vspace{-3mm}
In this setting, we consider arrays of two-level atoms with ground state $\ket{g}$ and Rydberg state $\ket{r}$, arranged into multiple clusters, i.e., collections of neighboring (mutually blockaded) atoms. We assume clusters far enough from each other to neglect their mutual interaction.
When the system is globally driven by a laser tuned to the $\ket{g}\leftrightarrow\ket{r}$ transition, the system's Hamiltonian reads:
\begin{equation} \label{eq:hamiltonian_gr}
    H=\frac{\Omega(t)}{2}\sum_{j}\left(e^{i\xi(t)}\ket{g_j}\bra{r_j}+{\rm h.c.}\right)+\sum_{\langle j,m\rangle}V_{jm}\hat{\mu}_j\hat{\mu}_m,
\end{equation}
with the second sum running only over pairs of blockaded atoms, i.e., atoms belonging to the same cluster; and where the same notation as in Eq.~\eqref{eq:hamiltonian} has been used.

Our study of this setting is notably motivated by the proposal in Ref.~\cite{cesaUnivQCGlobal}, which allows for universal quantum computation with global driving fields only, eliminating the need for local control. Additionally, such configurations are central to quantum simulation schemes based on reconfigurable atom arrays \cite{maskara2023programmable}.

In the blockade limit ($V_{jm}\gg \Omega(t)$ for all blockaded pairs), similarly to the gate setting, multiple Rydberg excitations within the same cluster are prohibited. Consequently, the dynamics of a single cluster simplifies to that of an effective TLS, described by Eq.~\eqref{eq:Hks}. A cluster of $n_k$ atoms evolves in the space spanned by its effective ground state ${\ket{g,k}=\ket{g}^{\otimes n_k}}$ and effective excited state ${\ket{e,k}=\tfrac{1}{\sqrt{n_k}}\sum_{l=1}^{n_k}\ket{g}^{\otimes (l-1)}\otimes\ket{r}\otimes\ket{g}^{\otimes (n_k-l)}}$, with effective Rabi frequency $\sqrt{n_k}\,\Omega$. Since clusters with the same number $n_k$ of atoms undergo the same dynamics, it is sufficient to distinguish between clusters with different numbers of atoms. Indexing these clusters by $k\in\left\{1,\dots,N\right\}$, we express the state of the $k$-th cluster as ${\ket{\psi_k}=\psi_{g,k}\ket{g,k}+\psi_{e,k}\ket{e,k}}$ and, from the Schr\"odinger equation, it follows that $\psi_{g,k}$ and $\psi_{e,k}$ evolve according to Eq.~\eqref{eq:Schrodinger}.

In the cluster setting, we are interested in performing a given unitary operation on each effective TLS, i.e., on each set of clusters with the same number of atoms, in parallel. In many relevant cases, including Ref.~\cite{cesaUnivQCGlobal}, it is actually sufficient to consider the action of the unitaries on the ground states of the effective TLSs. In this way, the task reduces to simultaneously evolving the TLSs from their ground states to given target states $\ket{\psi_k}$, i.e., $\ket{g,k}\rightarrow\ket{\psi_k}$. For instance, a process of interest in the following is the simultaneous excitation process, consisting in bringing two effective TLSs from their ground to their excited state: $\ket{g,1}\rightarrow\ket{e,1}$, $\ket{g,2}\rightarrow\ket{e,2}$. Other processes of interest in this setting are the selective rotation and selective phase processes, which are detailed in Sec.~\ref{Sec:results}.

In summary, the gate and cluster settings are both described by the same mathematical formalism, with dynamics governed by Eq.~\eqref{eq:Schrodinger}. In the following section, we thus cast time-optimal control problems for both settings into a unified HJB framework, and solve them for various target processes of interest in each setting. 

\section{Method: the HJB equation}\label{sec:HJB}
\vspace{-3mm}
In this section, we show how to apply the Bellman formalism to the time-optimal control of systems whose evolution can be reduced to that in Eq.~\eqref{eq:Schrodinger}. We derive the associated Hamilton-Jacobi-Bellman equations and discuss how to compute their solution through a generalized method of characteristics.
\vspace{-3mm}
\subsection{Main principles}
\vspace{-3mm}
The Bellman principle of optimality states that every part of an optimal trajectory is itself optimal for the associated optimization problem \cite{bellman1957dynamic,kirkOptimalControlTheory}: given three points $a$, $b$, and $c$ in the state space, if $a$-$b$-$c$ is the optimal path from $a$ to $c$, then $b$-$c$ is the optimal path from $b$ to $c$.
This principle allows to decompose an optimization problem into a sequence of smaller optimization instances, leading to an algorithmic procedure known as dynamic programming. In the case of continuous-time dynamical systems, this results in a partial differential equation for the optimal cost on the state space, known as the \textit{Hamilton-Jacobi-Bellman} (HJB) equation.
\vspace{-3mm}
\subsubsection{Background on the HJB formalism}
\vspace{-3mm}
For completeness, we start by presenting the basic concepts underlying the HJB approach to optimal control. We
consider a controlled dynamical system, with (real) state variables $\vec{x}(\tau)$, evolving according to the following equations of motion: ${\dot{\vec{x}}(\tau)=\vec{a}(\vec{x}(\tau),\vec{u}(\tau),\tau)}$, with $\tau\in[t,T]$. Here, $\vec{u}(\tau)$ represents the set of controls acting on the system and the dot denotes the derivative with respect to $\tau$. We are interested in controlling this evolution in order to minimize an assigned cost functional of the type:
\begin{align}\label{eq:cost_function}
J(\vec{x}(t),t,\vec{u}(\tau))=h(\vec{x}(T),T)+\int_{t}^{T}g(\vec{x}(\tau),\vec{u}(\tau),\tau)d\tau,
\end{align}
where $h$ and $g$ are given functions specifying the cost. Since $\vec{x}(\tau)$ can be computed from the equations of motion, $J$ is uniquely specified by the initial time $t$ of the dynamics, the initial state $\vec{x}(t)$, and the control's profile $\vec{u}(\tau)$ along the whole evolution. 

For any initial time $t$ and initial state $\vec{x}\equiv\vec{x}(t)$, we can define the optimal cost function $J^*(\vec{x}, t)$ as the optimum of $J$ over all possible controls, given these initial conditions. Applying the dynamic programming method to this continuous-time optimal-control problem yields the following HJB equation for the optimal cost function \cite{kirkOptimalControlTheory}: \begin{align}\label{eq:HJB0}
    -\frac{\partial J^*(\vec{x},t)}{\partial t}=
    \min_{\vec{u}}\left\{g(\vec{x},\vec{u},t)+\vec{a}(\vec{x},\vec{u},t)\cdot\nabla_{\vec{x}}J^*(\vec{x},t)\right\}.
\end{align}
This equation is typically solved backward in time, starting from the final-time condition $J^*(\vec{x}, T) = h(\vec{x}, T)$. From Eq.~\eqref{eq:HJB0}, we get both the (globally) optimal cost function $J^*(\vec{x},t)$, solution of the equation, and the optimal control function over the state space $\vec{u}^*(\vec{x},t)$, which minimizes the argument on the right-hand side. The function $\vec{u}^*(\vec{x},t)$ gives the control to choose to follow the optimal path when the system is in $\vec{x}$ at time $t$, and it allows us to reconstruct the optimal trajectories directly from the dynamics.
For example, the optimal trajectory $\vec{x}^*(t)$ passing through $\vec{x}_1$ at time $t_1$ solves the following equation: $\dot{\vec{x}}^*(t)=\vec{a}(\vec{x}^*(t),\vec{u}^*(\vec{x}^*(t),t),t)$, with $\vec{x}^*(t_1)=\vec{x}_1$. 

While the setting above is rather general, in the following we are interested in simplified problems where both the dynamics (i.e., $\vec{a}$) and the cost (i.e., $h$ and $g$) do not depend explicitly on time. In this case, Eq.~\eqref{eq:HJB0} admits a time-independent solution, which can be computed from the associated stationary HJB equation. In particular, we consider the time-optimal control problem of determining the minimum time needed to reach a target set $\Sigma$ of the state space, from an arbitrary point $\vec{x}$. This problem can be formulated by choosing $J=T-t$ as cost function (i.e., setting $h=0$ and $g=1$ in Eq.~\eqref{eq:cost_function}) and imposing that it vanish on the target set. Thus, from Eq.~\eqref{eq:HJB0}, we obtain the following stationary HJB equation:
\begin{equation}\label{eq:HJBs}
   \min_{\vec{u}}\left\{\vec{a}(\vec{x},\vec{u})\cdot\nabla_{\vec{x}}J^*(\vec{x})\right\}=-1,\:\:\:\text{with }J^*\rvert_{\Sigma}=0.
\end{equation}
 
\vspace{-3mm}
\subsubsection{HJB equations for globally driven Rydberg processes}
\vspace{-3mm}
We now apply the above formalism to the $N$-TLS model presented in Sec.~\ref{sec:model}. For this, we identify the state vector $\vec{x}$ with $\vec{\psi}=\left(\psi_{g,1},\psi_{e,1},...,\psi_{g,N},\psi_{e,N}\right)$ and the control $\vec{u}(t)$ with $(\Omega(t),\xi(t))$ (with $\Omega(t)$ bounded); the dynamics $\vec{a}$ is given by Eq.~\eqref{eq:Schrodinger}. The stationary HJB equation thus reads:
\begin{equation}
\label{eq:HJBscPSI}
   \min_{\xi,\Omega}\left\{\dot{\vec{\psi}}\cdot\nabla_{\vec{\psi}}J^*+\dot{\overline{\vec{\psi}}}\cdot\nabla_{\overline{\vec{\psi}}}J^*\right\}=-1,
\end{equation}
where we account for the complex nature of the state vector $\vec{\psi}$ by introducing its complex conjugate vector $\overline{\vec{\psi}}$.
The minimization over the controls is simplified by observing that $\Omega(t)$ appears as a prefactor in the quantity to minimize (see Eqs.~(\ref{eq:Schrodinger},\ref{eq:HJBscPSI})), whose sign can be arbitrarily flipped via the transformation $\xi\rightarrow\xi+\pi$. 
Consequently, the optimum in Eq.~\eqref{eq:HJBscPSI} is achieved when $\Omega(t)$ assumes its maximum value at all times ($\Omega(t)\equiv\Omega$), leaving the laser phase $\xi(t)$ as the only control parameter. 

For $N$ independent TLSs, Eq.~\eqref{eq:HJBscPSI} depends on the $2N$ complex (i.e., $4N$ real) variables in $\vec{\psi}$. However, exploiting the normalization of the TLSs' states, it is possible to express it as a function of $3N$ real variables $\vec{x}={\left(\theta_1,\dots,\theta_N,\phi_1,\dots,\phi_N,\phi_{g,1},\dots,\phi_{g,N}\right)}$ by means of the following change of variables:
\begin{align}\label{eq:Bloch}
\begin{split}
    \psi_{e,k}&={\rm cos}(\theta_k/2)e^{i(\phi_{g,k}-\phi_k)},\\
    \psi_{g,k}&={\rm sin}(\theta_k/2)e^{i\phi_{g,k}}.
\end{split}
\end{align}
Here, $\theta_k$ and $\phi_k$ geometrically correspond to the latitude and the longitude of the $k$-th TLS's state on the Bloch sphere, respectively, and $\phi_{g,k}$ to the phase of its ground state coefficient. Via this change of variables, the minimization in Eq.~\eqref{eq:HJBscPSI} can be performed analytically, giving the following expression of the optimal control (see Appendix~\ref{App:HJB}):
\begin{multline}\label{eq:optictrl}
    \xi^*(\vec{x})={\rm arg}\left\{\sum_{k=1}^N\left[-i\sqrt{n_k}\,e^{i\phi_k}\frac{\partial J^*}{\partial \theta_k}+\frac{\sqrt{n_k}\,e^{i\phi_k}}{{\rm tan}(\theta_k)}\frac{\partial J^*}{\partial \phi_k}\right.\right.\\
    \left.\left.+\frac{\sqrt{n_k}\,e^{i\phi_k}}{2{\rm tan}(\theta_k/2)}\frac{\partial J^*}{\partial \phi_{g,k}}
    \right]\right\},
\end{multline}
and the following expression of the  HJB equation:
\begin{multline}\label{eq:HJBgeneral}
    1=\Omega\left\lvert \sum_{k=1}^N\left[i\sqrt{n_k}\,e^{i\phi_k}\frac{\partial J^*}{\partial \theta_k}-\frac{\sqrt{n_k}\,e^{i\phi_k}}{{\rm tan}(\theta_k)}\frac{\partial J^*}{\partial \phi_k}\right.\right.\\
    \left.\left.-\frac{\sqrt{n_k}\,e^{i\phi_k}}{2{\rm tan}(\theta_k/2)}\frac{\partial J^*}{\partial \phi_{g,k}}
    \right]
    \right\rvert.
\end{multline}
The knowledge of $J^*(\vec{x})$, computed by solving Eq.~\eqref{eq:HJBgeneral} with the boundary condition $J^*\rvert_{\Sigma}=0$, contains all the information about the solution of the optimal control problem. Specifically, the optimal trajectories can be directly derived from the optimal laser phase $\xi^*(\vec{x})$, given by Eq.~\eqref{eq:optictrl}. We last note that, depending on the target process, Eq.~\eqref{eq:HJBgeneral} can be further simplified by eliminating irrelevant variables (see Sec.~\ref{sec:2qubit}).
\vspace{-3mm}
\subsection{Generalized method of characteristics}\label{sec:characteristics}
\vspace{-3mm}
Due to its nonlinear nature, Eq.~\eqref{eq:HJBgeneral} does not admit a classical (i.e., smooth) solution. It however admits a unique generalized \textit{viscosity solution} \cite{crandallViscositySolutions,crandallPropertiesViscosity} corresponding to the global optimum of the optimization problem, which can be computed by means of a generalization of the method of characteristics \cite{subbotinMinimax,subbotinaHJCharacs}.

The method of characteristics turns a first-order partial differential equation (PDE) into a system of coupled ordinary differential equations (ODEs), which governs the evolution of the solution along parametric curves known as characteristics. For a HJB equation of the type ${F(\vec{x},\vec{p},J^*)=0}$, with $\vec{p}\equiv\nabla_{\vec{x}}J^*$, the characteristics are defined as the solution of the following ODE system: $\dot{\vec{x}}(s)=\nabla_{\vec{p}}F$, $\dot{\vec{p}}(s)=-\nabla_{\vec{x}}F-(\partial F/\partial J^*)\vec{p}$, and $\dot{J}^*(s)=\vec{p}\cdot\nabla_{\vec{p}}F$ (where the dot denotes the derivative with respect to the curve parameter $s$). Integrating the characteristic equations, with appropriate initial conditions, allows to rebuild the solution of the original equation \cite{evansPDE}. However, when the characteristics cross in the state space, this method fails, returning multiple values for $J^*(\vec{x})$. The crossing regions, known as shocks, reflect the absence of a smooth solution of the HJB equation, which indeed only admits a generalized solution (e.g., continuous but not differentiable). 

Nevertheless, this generalized solution still depends only on the values of $J^*$ along the characteristics \cite{sethianOrderedUpwind} and thus can be rebuilt through a generalization of the method. The generalized solution can be obtained by taking, at each point, the smallest of the values of $J^*$ returned by the characteristics passing through the point. We refer to the characteristic corresponding to this minimum value, for a given point, as the optimal one. 
This construction gives the so-called minimax solution of the HJB equation, a concept equivalent to the viscosity one \cite{subbotinaHJCharacs,subbotinMinimax,clarkeNonsmooth,bayenMinimax}. 

The initial conditions for the integration of the characteristics are derived from the boundary condition of the HJB equation, i.e., the target of the optimal control problem. For a state space of dimension $M$, the set of characteristics is indexed by $M-1$ free parameters in the choice of the initial conditions of the characteristic equations (see Appendix~\ref{App:characsmethod}).

In the case of Eq.~\eqref{eq:HJBgeneral}, the characteristics correspond to the physical trajectories of the system and the parameter $s$ can simply be interpreted as the time variable along them ($J^*(s)=s$). Thus, finding the optimal characteristic for a given state space point $\vec{x}$ corresponds to finding the globally optimal physical trajectory from $\vec{x}$ to the target set $\Sigma$. The associated optimal control $\xi^*(s)$ is then directly computed from Eq.~\eqref{eq:optictrl}. Furthermore, this method can also return analytical necessary conditions that the control should satisfy for optimality, such as its initial slope (see Appendix~\ref{App:characsmethod}).

\vspace{-3mm}
\subsection{A toy example: the single-qubit case}\label{Sec:1qubit}
\vspace{-3mm}
\begin{figure}
    \centering
    \includegraphics[width=\linewidth]{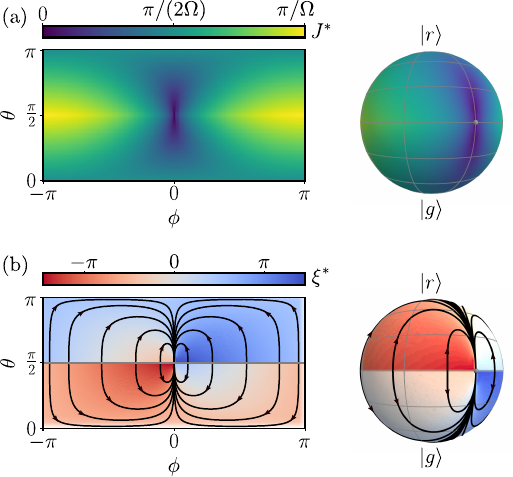}
    \caption{Solution of the time-optimal control problem for a single driven qubit with target point $(\theta_0,\phi_0)=(\pi/2,0)$ on the equator. (a) Left: Solution $J^*(\theta,\phi)$ of the HJB equation in Eq.~\eqref{eq:HJBsinglequbit}. Right: Representation of  $J^*(\theta,\phi)$ on the qubit's Bloch sphere (target point in gray). The function is symmetric around the shock line situated on the equator, where it is continuous but not differentiable. (b) Left: Optimal control values $\xi^*(\theta,\phi)$, i.e., the optimal driving phase to go towards the target point $(\theta_0,\phi_0)$ from the point $(\theta,\phi)$, computed via Eq.~\eqref{eq:optictrl}. The corresponding optimal trajectories, i.e., (reversed) characteristics of the HJB equation in Eq.~\eqref{eq:HJBsinglequbit}, are represented as oriented black lines. Right: Representation of $\xi^*(\theta,\phi)$ on the qubit's Bloch sphere (target point in gray). The shock line is here visible in the $\pi$ jump of the optimal laser phase.}
    \label{fig:1qubit}
\end{figure}
In order to illustrate the above approach on a simple example, we start by considering the case of a single driven qubit (i.e., a single TLS). We apply the HJB formalism (Eq.~\eqref{eq:HJBgeneral} with $N=1$) to solve the associated time-optimal control problem. Here, the $\phi_g$ variable represents an irrelevant global phase and thus it is physically meaningful to consider boundary conditions independent of it: $\partial J^*/\partial\phi_g=0$ for $\vec{x}\in\Sigma$. Furthermore, observing that the HJB equation itself does not depend explicitly on $\phi_g$, it follows that $\partial J^*/\partial\phi_g$ is constant and can be set to zero directly in the original equation, which thus simplifies to:
\begin{align}\label{eq:HJBsinglequbit}
    1=\Omega\left\lvert i\frac{\partial J^*}{\partial \theta}-\frac{1}{{\rm tan}(\theta)}\frac{\partial J^*}{\partial \phi}
    \right\rvert.
\end{align}

The simplest problem we can consider is to determine the minimum time needed to reach a certain latitude $\theta_0$ on the Bloch sphere, from another arbitrary one. In this case, Eq.~\eqref{eq:HJBsinglequbit} further simplifies: the boundary condition (as the equation itself) does not depend explicitly on $\phi$, and so $\partial J^*/\partial \phi$ can be set to zero from the beginning. We obtain $1=\Omega\left\lvert\partial J^*/\partial\theta\right\rvert$, which is solved by ${J^*(\theta)=\lvert\theta-\theta_0\rvert/\Omega}$. This corresponds to driving the system along the meridians of the Bloch sphere.

A slightly more general problem consists in computing the minimum time to reach a target point $(\theta_0,\phi_0)$ on the Bloch sphere from any other point $(\theta,\phi)$. For this, we must solve Eq.~\eqref{eq:HJBsinglequbit} with the boundary condition $J^*(\theta_0,\phi_0)=0$. Since the instantaneous rotation axis is constrained to lie in the equatorial plane (see Eq.~\eqref{eq:Schrodinger}), the optimal trajectories are not the sphere geodesics passing through the target. The solution of the HJB equation is nontrivial, but it admits a simple analytical expression in the neighborhood of the target point: ${J^*(\theta,\phi)\simeq\sqrt{(\theta-\theta_0)^2+\tan^2(\theta_0)(\phi-\phi_0)^2}}$. Here, the level sets of $J^*$ are (to leading order) ellipses centered at $(\theta_0,\phi_0)$.
In the whole space, the equation can be solved by resorting to the generalization of the method of characteristics (see Sec.~\ref{sec:characteristics}). The characteristic equations associated to Eq.~\eqref{eq:HJBsinglequbit} read:

\vspace{-5mm}
\begin{align}\label{eq:characs1qubit}
    \begin{aligned}
        \dot{\theta}&=\Omega^2 p_{\theta}, & \qquad \dot{p}_{\theta}&=\Omega^2 \frac{p_{\phi}^2}{{\rm tan}(\theta){\rm sin}^2(\theta)},\\
        \dot{\phi}&=\Omega^2 \frac{p_{\phi}}{{\rm tan}^2(\theta)}, & \dot{p}_{\phi}&=0, 
    \end{aligned}
\end{align}
where $p_{\theta}=\partial J^*/\partial\theta$ and $p_{\phi}=\partial J^*/\partial\phi$. 
The boundary condition for the cost function fixes the initial condition on the state variables to be $\theta(0)=\theta_0$ and $\phi(0)=\phi_0$. The set of characteristics is then generated by varying the initial conditions on the momentum variable, $p_{\theta}(0)=p_{\theta,0}$ and $p_{\phi}(0)=p_{\phi,0}$, constrained by ${p_{\theta,0}^2+p_{\phi,0}^2{\rm cot}^2(\theta_0)=1/\Omega^2}$, due to the HJB equation at the target. 

The characteristic equations can be solved analytically (see Appendix~\ref{App:analytic}) and, in the case of $\theta_0=\pi/2$ and $\phi_0=0$ (i.e., for a target point on the equator), we obtain:
\begin{align} \label{eq:characteristicsEquator}
\begin{split}
    &\theta(s)=\arccos\left[\frac{-\sin\left(\pm \Omega s A\right)}{A}\right],\\
    &\phi(s)= -\Omega^2 p_{\phi} s+\arctan\left[\frac{\Omega p_{\phi} \tan\left(\Omega s A\right)}{A}\right],
    \end{split}
\end{align}
where $A=\sqrt{1+(\Omega p_{\phi})^2}$. 
In addition, the value of the optimal control $\xi^*(s)$ can be computed along these trajectories via Eq.~\eqref{eq:optictrl}, returning $\xi^*(s)=-\Omega^2 p_{\phi}s\pm\pi/2$. 
We find two branches of characteristics corresponding to the $\pm$ sign in Eq.~\eqref{eq:characteristicsEquator}, whose inversion returns two multivalued functions, namely $s_{\pm}(\theta,\phi)$. The solution of the HJB equation can be then constructed as ${J^*(\theta,\phi)=\min\left\{s_+(\theta,\phi),s_-(\theta,\phi)\right\}}$, where the minimum is taken over all the values assumed by $s_+$ and $s_-$. The two branches cross at the equatorial line (i.e., $\theta=\pi/2$), which so appears to be a shock line for the HJB equation, where the derivative of $J^*$ 
becomes discontinuous. In Fig.~\ref{fig:1qubit}, we show $J^*(\theta,\phi)$ on the $(\theta,\phi)$ space, i.e., on the Bloch sphere, as well as the optimal control $\xi^*(\theta,\phi)$ computed via Eq.~\eqref{eq:optictrl}, and the corresponding optimal trajectories. The presence of the shock is also clearly visible in the discontinuity of the control at the equator.

\section{Two two-level systems}\label{sec:2qubit}
\vspace{-3mm}
In this section, we focus on time-optimal control problems for two effective independent two-level systems. This is the main scenario of interest, which describes both two-qubit gates and the universal quantum computing scheme of Ref.~\cite{cesaUnivQCGlobal}. Accordingly, we consider the HJB equation in Eq.~\eqref{eq:HJBgeneral} with $N=2$. Unless otherwise specified, we set $n_1=1$ and $n_2=2$ (extensions to other values of $n_1$ and $n_2$ are straightforward).

\vspace{-3mm}
\subsection{HJB equation for two TLSs}\label{sec:HJB2tls}
\vspace{-3mm}
We now simplify Eq.~\eqref{eq:HJBgeneral} to a minimum-dimensional form, suitable to address most relevant target processes in both gate and cluster settings. We discuss how to overcome the challenges posed by the singularity of the HJB equation through an ad-hoc, physics-informed tailoring of the initial conditions of the characteristic equations.

\vspace{-3mm}
\subsubsection{Simplified HJB equations}\label{sec:HJBeqs2tls}
\vspace{-3mm}
In the two-TLS case, the state variables in Eq.~\eqref{eq:HJBgeneral} are $(\theta_1,\theta_2,\phi_1,\phi_2,\phi_{g,1},\phi_{g,2})$. However, for most relevant processes, this equation can be further simplified, thereby reducing the dimensionality. Indeed, if a given variable does not explicitly appear in the HJB equation, and the boundary condition does not depend on it, it can be eliminated from the equation by setting the corresponding partial derivative to zero.

With this goal in mind, we observe that $e^{i\phi_1}$ can be factored out of Eq.~\eqref{eq:HJBgeneral}. Thus, the equation does not depend explicitly on $\phi_1$ and $\phi_2$, but only on their difference $\phi=\phi_2-\phi_1$. If also the target is independent of $\phi_1$ and $\phi_2$, it is possible to eliminate one of these degrees of freedom. Analogously, observing that Eq.~\eqref{eq:HJBgeneral} does not depend explicitly on the ground state phases $\phi_{g,1}$ and $\phi_{g,2}$, these variables can be eliminated from the original equation when considering targets independent of them. In this case, Eq.~\eqref{eq:HJBgeneral} simplifies to (see Appendix~\ref{App:HJB2tls}):
\begin{align} \label{eq:HJBampli}
1&=\Omega\left\lvert i\frac{\partial J^*}{\partial \theta_1}+\sqrt{2}ie^{i\phi}\frac{\partial J^*}{\partial \theta_2}+\left(\frac{1}{{\rm tan}(\theta_1)}-\frac{\sqrt{2}e^{i\phi}}{{\rm tan}(\theta_2)}\right)\frac{\partial J^*}{\partial \phi} \right\rvert.
\end{align}
In the gate setting, it is however common to consider targets that explicitly depend on the ground state phases via the \textit{gate phase} variable, defined as $\Phi\equiv\phi_{g,2}-2\phi_{g,1}$. In this case, Eq.~\eqref{eq:HJBgeneral} simplifies to the following form:
\begin{align} \label{eq:HJBgates}
1&=\Omega\left\lvert i\frac{\partial J^*}{\partial \theta_1}+\left(\frac{1}{{\rm tan}(\theta_1)}-\frac{\sqrt{2}e^{i\phi}}{{\rm tan}(\theta_2)}\right)\frac{\partial J^*}{\partial \phi} +\sqrt{2}ie^{i\phi}\frac{\partial J^*}{\partial \theta_2}  \right.\notag\\
&\left.\qquad+ \left(\frac{1}{{\rm tan}(\theta_1/2)}-\frac{e^{i\phi}}{\sqrt{2}{\rm tan}(\theta_2/2)}\right)\frac{\partial J^*}{\partial \Phi}\right\rvert.
\end{align}

As discussed in Sec.~\ref{sec:gate} and~\ref{sec:cluster}, both in the gate and in the cluster settings, we are interested in processes consisting in bringing both TLSs from their ground state to a given target state. The point corresponding to having both TLSs in their ground state is ${\vec{x}_0\equiv(\theta_{1,0},\theta_{2,0},\phi_0,\Phi_0)=(\pi,\pi,0,0)}$ (where the fourth component is understood to be omitted in the case of Eq.~\eqref{eq:HJBampli}), while the target point, denoted by $\vec{x}_t$, depends on the process under study. In the HJB formalism, this translates, in principle, to solving Eqs.~(\ref{eq:HJBampli},\ref{eq:HJBgates}) with the boundary condition $J^*(\vec{x}_t)=0$, and computing $J^*(\vec{x}_0)$. However, since all the processes share the same starting point $\vec{x}_0$, and the dynamics in Eq.~\eqref{eq:Schrodinger} is invertible, it is more convenient to consider the reversed processes, i.e., going from $\vec{x}_t$ to $\vec{x}_0$. We thus set $J^*(\vec{x}_0)=0$ as boundary condition and compute $J^*(\vec{x}_t)$. This choice indeed leads to equivalent results but allows us to reduce all optimal control problems of interest to the solution of Eqs.~(\ref{eq:HJBampli},\ref{eq:HJBgates}) with a unique boundary condition, given by $J^*(\vec{x}_0)=0$.

\vspace{-3mm}
\subsubsection{Singularities and characteristics}\label{Sec:characs2q}
\vspace{-3mm}
The characteristic equations of the above HJB equations (Eqs.~(\ref{eq:HJBampli},\ref{eq:HJBgates})) can be analytically derived, and are given explicitly in Appendix~\ref{App:characs}. Then, to approach the optimal control problem via the generalized method of characteristics, we must in principle integrate the characteristic equations (with initial conditions $\vec{x}(0)=\vec{x}_0$ and $\vec{p}(0)$ free), and deal with shocks as outlined in Sec.~\ref{sec:characteristics}. However, we find that the characteristic equations exhibit singular coefficients at $\vec{x}_0$, preventing straightforward numerical integration. To overcome this, we devise a physics-informed integration strategy.

Since the characteristics correspond to physical trajectories, they must satisfy \textit{short-time} constraints given by the equations of motion in Eq.~\eqref{eq:Schrodinger}. In Appendix~\ref{App:shorttime}, we present a detailed analysis of Eq.~\eqref{eq:Schrodinger}, which provides us with the short-time expansions of the state variables. In this way, we first of all show that, although $\phi$ is not properly defined when both TLSs are in the ground state, since on both TLSs the same driving phase $\xi$ acts, it must tend to zero at the start (precisely, $\phi(s)=O(s^3)$). Second, since the Rabi frequencies acting on the TLSs are $\Omega$ and $\sqrt{2}\,\Omega$, it follows that $\theta_1\sim\Omega s$ and $\theta_2\sim\sqrt{2}\,\Omega s$ at short time. 

These expressions greatly simplify the characteristic equations at the start, eliminate some of the singularities, and allow us to deal with the others. In particular, since $\dot{p}_{\phi}=p_{\phi}/s+O(s)$, it results that $p_{\phi}(s)\sim f_{\phi,0} s$ at short time, where $f_{\phi,0}$ is 
a free parameter (representing the initial derivative of $p_{\phi}$). The characteristics are thus parameterized by either of $p_{\theta_1,0}$ and $p_{\theta_2,0}$ (with the other being constrained by the HJB equation), $f_{\phi,0}$, and $p_{\Phi,0}$ (in the case of Eq.~\eqref{eq:HJBgates}).

\vspace{-3mm}
\subsection{Results}\label{Sec:results}
\vspace{-3mm}
To find the optimal time to reach the point $\vec{x}_0$ from the target point $\vec{x}_t$, we numerically integrate the characteristics starting from $\vec{x}_0$, indexed by $(p_{\theta_1,0},f_{\phi,0})$ or $(p_{\theta_1,0},f_{\phi,0},p_{\Phi,0})$, and find the one(s) that cross $\vec{x}_t$. We perform this search by minimizing the distance between the characteristics and $\vec{x}_t$, over the 2- or 3-dimensional parameter space. Eventually, the fastest resulting trajectory corresponds to the global optimum, and the corresponding optimal control $\xi^*$ can be computed from Eq.~\eqref{eq:optictrl} (performing the same reductions as in Sec.~\ref{sec:HJB2tls}). We note that, employing the short-time constraints described in the previous subsection and in Appendix~\ref{App:shorttime}, one can show that for all characteristics of Eqs.~(\ref{eq:HJBampli},\ref{eq:HJBgates}), the time-derivative of $\xi^*$ vanishes at the start.  
Thus, optimal pulses starting from (ending at) $\vec{x}_0$ must be resonant (i.e., $\dot{\xi}^*=0$) at the start (end), which is indeed seen in the results below. 

We now present a series of results concerning key processes for various quantum computation schemes with neutral atoms, in both gate and cluster settings. We note that the numerical values given in this section for $p_{\theta_1,0}$ and $p_{\Phi,0}$ are in units of $1/\Omega$.

\begin{figure}
    \centering
    \includegraphics[width=\linewidth]{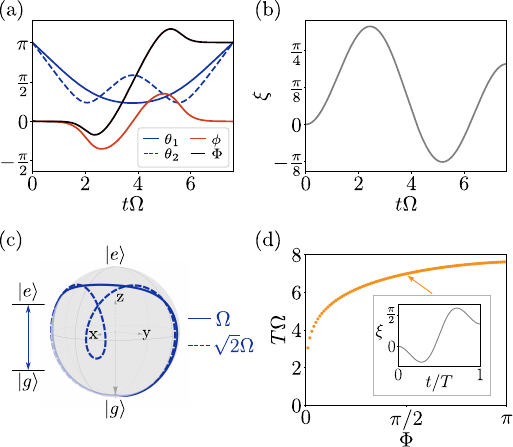}
    \caption{Optimal control of two-qubit controlled-phase (C-$\Phi$) gates. (a) Optimal trajectory of the CZ gate ($\Phi=\pi$), obtained from the characteristic equations of Eq.~\eqref{eq:HJBgates} (given in Appendix~\ref{App:characs}). This trajectory corresponds to choosing $(p_{\theta_1,0},f_{\phi,0},p_{\Phi,0})=(-0.250,0.888,-0.131)$ at the start. (b) Optimal control profile associated to this trajectory, reconstructed via Eq.~\eqref{eq:optictrl}. This result is analogous to that previously obtained in Ref.~\cite{janduraTimeOptimalTwoThreeQubit2022}. (c) Bloch sphere representation of the trajectories of the two effective TLSs during the CZ gate's execution. (d) Optimal time to realize arbitrary controlled-$\Phi$ gate, as a function of $\Phi$. The values are in agreement with those obtained via an approximate ansatz in Ref.~\cite{everedHighFidelityParallel}. Inset: Optimal control pulse for the C-($\pi/2$) gate.}
    \label{fig:cz}
\end{figure}

\paragraph{{\bf Controlled-phase gates.}} A central process involving two effective TLSs is 
the controlled-Z (CZ) gate \cite{levineParallelGates}, which leaves all two-qubit computational basis states unchanged, except for the $\ket{1,1}$ state which acquires a minus sign. In the computational subspace, the CZ gate acts as ${\rm CZ}\ket{z_1,z_2}=(-1)^{z_1z_2}\ket{z_1,z_2}$ (with ${z_1,z_2\in\{0,1\}}$). In the effective TLS formalism, it corresponds (up to single-qubit rotations) to both effective two-level systems returning to their ground states with phases $\phi_{1,g}$ and $\phi_{2,g}$ such that $\Phi=\phi_{2,g}-2\phi_{1,g}=\pi$, i.e.,
\begin{align}
    \begin{split}
        \ket{g,1}&\rightarrow e^{i\alpha}\ket{g,1},\\
        \ket{g,2}&\rightarrow e^{i(2\alpha+\pi)}\ket{g,2},
    \end{split}
\end{align}
with $\alpha$ free. Here, thus, the relevant HJB equation is Eq.~\eqref{eq:HJBgates}, and the target point for this process is ${\vec{x}_t=(\theta_1,\theta_2,\phi,\Phi)=(\pi,\pi,0,\pi)}$. We compute the characteristics and find that the fastest one to reach $\vec{x}_t$ corresponds to the choice ${(p_{\theta_1,0},f_{\phi,0},p_{\Phi,0})=(-0.250,0.888,-0.131)}$. The corresponding protocol has a duration of $T=7.611/\Omega$ and corresponds to that first proposed by Jandura and Pupillo in Ref.~\cite{janduraTimeOptimalTwoThreeQubit2022}. In Fig.~\ref{fig:cz}, we show the associated trajectory and optimal control. We note that the control can be interpreted as a smoothed version of a three-pulse sequence, consisting of two pulses with same detuning ($\dot{\xi}=\Delta_1$) and duration separated by another pulse with opposite detuning ($\dot{\xi}=-\Delta_2$); that sequence realizes an exact gate with a duration less than 1\% longer than the optimal time. 

In addition to the canonical CZ gate, we consider a continuous family of controlled-phase (C-$\Phi$) gates, parameterized by the phase $\Phi$ acquired by the $\ket{1,1}$ state: ${\rm C\Phi}\ket{z_1,z_2}=e^{i \Phi z_1z_2}\ket{z_1,z_2}$, up to single-qubit rotations. The CZ gate corresponds to the special case $\Phi=\pi$. To find optimal times and controls for all these gates, while keeping $\vec{x}_0=(\pi,\pi,0,0)$ fixed, we set $\vec{x}_t=(\pi,\pi,0,\Phi)$ and perform the search for different values of $\Phi$. In Fig.~\ref{fig:cz}, we plot the optimal times for this family of gates as a function of the gate phase $\Phi$. The optimal time is an increasing function of $\Phi$ (the $\Phi=0$ gate is the identity, which can trivially be achieved in zero time). The resulting optimal times are in good agreement with those found by means of a variational ansatz for the pulse shape in Ref.~\cite{everedHighFidelityParallel}.

\begin{figure}
    \centering
    \includegraphics[width=\linewidth]{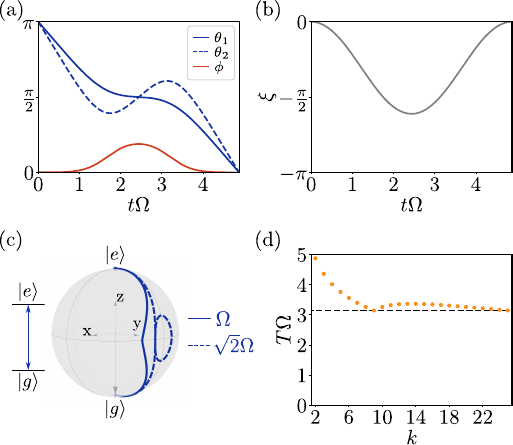}
    \caption{Optimal control of the simultaneous excitation process: on the Bloch sphere, the two effective TLSs evolve from their south pole $(\theta_1,\theta_2)=(\pi,\pi)$ to their north pole ${(\theta_1,\theta_2)=(0,0)}$. (a) Optimal trajectory obtained from the characteristic equations of Eq.~\eqref{eq:HJBampli} (given in Appendix~\ref{App:characs}). This trajectory corresponds to choosing ${(p_{\theta_1,0},f_{\phi,0})=(-0.071,-1.532)}$ at the start. (b) Optimal control profile associated to this trajectory, reconstructed via Eq.~\eqref{eq:optictrl}. (c) Bloch sphere representation of the trajectories of the two effective TLSs during the process. (d) Analysis of the simultaneous excitation process in the general case of effective TLSs with Rabi frequencies $\Omega$ and $\sqrt{k}\,\Omega$: optimal time as a function of $k$.}
    \label{fig:downup}
\end{figure}

\paragraph{\bf Simultaneous excitation process.} 
We now consider the \textit{simultaneous excitation} process, consisting in driving both effective TLSs from ground to excited state (i.e., from south to north pole of the corresponding Bloch spheres). This is indeed the fundamental building block of the proposal of Ref.~\cite{cesaUnivQCGlobal}, and it can be used in the gate setting to design robust entangling gates \cite{fromonteilProtocolsRobustness}. Explicitly, it reads:
\begin{align}
    \begin{split}
        \ket{g,1}&\rightarrow e^{i\alpha}\ket{e,1},\\
        \ket{g,2}&\rightarrow e^{i\beta}\ket{e,2},
    \end{split}
\end{align}
with $\alpha$ and $\beta$ free. 
Here, since the global phases acquired by the effective TLSs are irrelevant, we need to consider Eq.~\eqref{eq:HJBampli}, and compute its solution at the point 
$\vec{x}_t=(0,0,0)$. Computing the characteristics, we find that the fastest one to reach $\vec{x}_t$ corresponds to the choice $(p_{\theta_1,0},f_{\phi,0})=(-0.071,-1.532)$. The duration of this protocol is $T=4.875/\Omega$, representing approximately a 10\% improvement over the implementation in Ref.~\cite{fromonteilProtocolsRobustness}. In Fig.~\ref{fig:downup}, we show the corresponding trajectory, as well as the associated optimal control. The optimal control pulse (see Fig.~\ref{fig:downup}(b))
can again be interpreted as a smoothed version of a simple pulse sequence, consisting of two pulses with opposite (constant) detuning, i.e., $\dot{\xi}=\pm \Delta$; this near-optimal implementation again takes less than 1\% longer than the optimal one.

Instead of $\Omega$ and $\sqrt{2}\,\Omega$, one can, in the cluster setting, consider two effective TLSs with Rabi frequencies of  $\Omega$ and $\sqrt{k}\,\Omega$ (i.e., $n_1=1$ and $n_2=k$): this corresponds to using clusters with a number $k>2$ of atoms. Although increasing the number of atoms in a blockade radius might be experimentally challenging, we see that increasing $k$ reduces the duration needed to perform the same process. To illustrate this, we compute the optimal controls for the above simultaneous excitation process, for different values of $k$, and show the results in Fig.~\ref{fig:downup}(d). The optimal time approaches $\pi/\Omega$ as $k$ increases, which is indeed the expected lower bound: there is no faster way for the first TLS to go from ground to excited state. Additionally, we notice that this lower bound is reached when $\sqrt{k}$ is odd. Indeed, in this case, the Rabi frequency enhancement is such that driving the first TLS with a $\pi$ pulse exactly results in the other TLS reaching its excited state as well. 

\paragraph{\bf Selective rotation processes.} In the quantum computation proposal of Ref.~\cite{cesaUnivQCGlobal}, the main ingredient for implementing effective single-qubit gates on encoded qubits is a process in which one of the two TLSs returns to its ground state after the evolution (thus following a closed-loop trajectory), while the other one is driven to some target latitude. We refer to this type of processes as \textit{selective rotations}. When the first TLS is the one that returns to its ground state, this process explicitly corresponds to:
\begin{align}
\label{eq:selective_rotations}
    \ket{g,1}&\rightarrow e^{i\alpha}\ket{g,1},\\
    \ket{g,2}&\rightarrow e^{i\beta}\sin(\theta_{2,\rm targ}/2)\ket{g,2}+e^{i\gamma}\cos(\theta_{2,\rm targ}/2)\ket{e,2},\nonumber   
\end{align}
with $(\alpha,\beta,\gamma)$ free and $\theta_{2,\rm targ}$ specifying the target. Conversely, when the second TLS returns to its ground state, the role of the TLSs in Eq.~\eqref{eq:selective_rotations} is exchanged and we denote the target latitude for the first TLS by $\theta_{1,\rm targ}$.

\begin{figure}
    \centering
    \includegraphics[width=\linewidth]{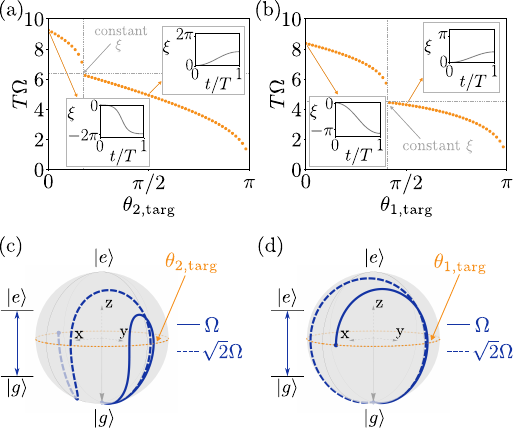}
    \caption{Optimal control for selective rotation processes: on the Bloch sphere, one of the two effective TLSs returns to its south pole, while the other is driven to a target latitude.
    (a) Optimal time to evolve from $(\theta_1,\theta_2)=(\pi,\pi)$ to $(\theta_1,\theta_2)=(\pi,\theta_{2,\rm targ})$, as a function of $\theta_{2,\rm targ}$. The discontinuity in the derivative of the curve corresponds to the resonant driving case, which results to be optimal, with time $2\pi/\Omega$. Insets: optimal drives for the cases of $\theta_{2,\rm targ}=0$ and $\theta_{2,\rm targ}=\pi/2$. (b) Conversely, optimal time to evolve from $(\theta_1,\theta_2)=(\pi,\pi)$ to $(\theta_1,\theta_2)=(\theta_{1,\rm targ},\pi)$, as a function of $\theta_{1,\rm targ}$. The discontinuity in the derivative of the curve corresponds to the resonant driving case, which results to be optimal, with time $2\pi/\sqrt{2}\Omega$. (c) Bloch sphere representation of the trajectories of the two effective TLSs corresponding to $\theta_{2, \rm targ}=\pi/2$. (d) Bloch sphere representation of the trajectories of the two effective TLSs corresponding to $\theta_{1, \rm targ}=\pi/2$.}
    \label{fig:thetas}
\end{figure}

As this process does not involve ground state phases, we integrate the characteristics of Eq.~\eqref{eq:HJBampli}, seeking the solution at the points $\vec{x}_t=(\pi,\theta_{2, \rm targ},\phi)$ and ${\vec{x}_t=(\theta_{1, \rm targ},\pi,\phi)}$. 
Here, the value of $\phi$ at the point $\vec{x}_t$ is left free. Indeed, since one of the two TLSs is in the ground state at this point, the longitude difference $\phi$ is undefined. We show the resulting optimal times as a function of $\theta_{2, \rm targ}$ or $\theta_{1, \rm targ}$ in Fig.~\ref{fig:thetas}.

Both plots of the optimal times present a discontinuity in their derivative, respectively at $\theta_{2, \rm targ}=(3-2\sqrt{2})\pi$ and at $\theta_{1, \rm targ}=(\sqrt{2}-1)\pi$. At these points, the optimal control corresponds to driving the system resonantly, i.e., with a constant laser phase. This feature evidences the property of the value function $J^*$ of being continuous but not everywhere differentiable. 
As discussed in Appendix~\ref{App:singlepulse}, to the right of the discontinuity, the associated optimal control can be interpreted as an improved version of a single-detuned-pulse implementation. To the left of the discontinuity, a single constant-detuning pulse is no longer enough to realize the process; however, a simple two- or three-pulse sequence can achieve this with a duration close to the optimal one.

In the specific cases of $\theta_{2,\rm targ}=0$ and $\theta_{1,\rm targ}=0$, we find that the optimal trajectories correspond, respectively, to the choices $(p_{\theta_1,0},f_{\phi,0})=(2.200, 0.061)$ and $(p_{\theta_1,0},f_{\phi,0})=(1.055,0.216)$ at the start; the associated optimal times are $T=9.244/\Omega$ and $T=8.364/\Omega$. The corresponding optimal controls are shown in Fig.~\ref{fig:thetas}. For $\theta_{2,\rm targ}=\pi/2$ and $\theta_{1,\rm targ}=\pi/2$, the optimal trajectories respectively correspond to the choices ${(p_{\theta_1,0},f_{\phi,0})=(0.955,0.546)}$ and ${(p_{\theta_1,0},f_{\phi,0})=(0.546,0.335)}$, and the associated optimal times are ${T=4.921/\Omega}$ and ${T=4.290/\Omega}$. The optimal trajectories and the corresponding controls are shown in Fig.~\ref{fig:thetas}.

\begin{figure}
    \centering
    \includegraphics[width=\linewidth]{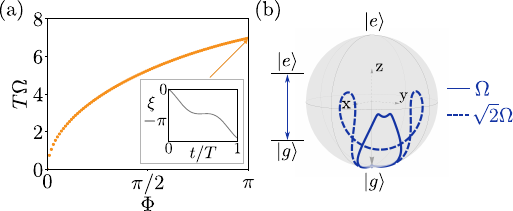}
    \caption{Optimal control for selective phase processes (i.e., effective controlled-phase gates in the setting of Ref.~\cite{cesaUnivQCGlobal}). Both effective TLSs return to their ground state, and the second one acquires a phase $\Phi$ (the phase acquired by the first TLS is irrelevant). (a) Optimal time to realize effective C-$\Phi$ gates in this setting, as a function of $\Phi$. Inset: optimal control pulse for the CZ ($\Phi=\pi$) case. (b) Bloch sphere trajectories of the two effective TLSs during the effective CZ gate's execution.}
    \label{fig:clustergates}
\end{figure}

\paragraph{\bf Selective phase processes.} Finally, the last process required in the universal quantum computation scheme of Ref.~\cite{cesaUnivQCGlobal} is the one that allows to realize effective controlled-phase gates between encoded qubits. In this case, the two effective TLSs should still go back to their ground states at the end of the process, but the gate phase $\Phi$ is now differently defined to be the phase acquired by the second TLS: $\Phi\equiv\phi_{g,2}$. We call this the \textit{selective phase} process; in terms of the TLSs' evolution, it corresponds to:
\begin{align}
    \begin{split}
        \ket{g,1}&\rightarrow e^{i\alpha}\ket{g,1},\\
        \ket{g,2}&\rightarrow e^{i\Phi}\ket{g,2},
    \end{split}
\end{align}
with $\alpha$ free (since the phase acquired by the first TLS is irrelevant) and $\Phi$ specifying the target. The variable $\phi_{g,1}$ can thus be eliminated in Eq.~\eqref{eq:HJBgeneral}, leading to a HJB equation slightly modified with respect to Eq.~\eqref{eq:HJBgates} (namely, the term involving $\tan(\theta_1/2)$ is removed). We note that this modification of the HJB equation also means that optimal pulses no longer have to start resonantly. Computing the characteristics of that equation, and performing the search for the point $\vec{x}_t=(\pi,\pi,0,\Phi)$ as above, 
yields the optimal time results shown in Fig.~\ref{fig:clustergates}. As in the standard controlled-phase gate case, the optimal times for arbitrary $\Phi$ again increase with $\Phi$. In the case of $\Phi=\pi$ (which corresponds to a CZ gate in the encoding of Ref.~\cite{cesaUnivQCGlobal}), we find an optimal protocol of duration $T=6.975/\Omega$. Once more, this protocol can be interpreted as a smoothed three-pulse sequence, with two detuned pulses separated by a resonant one.

\section{Three two-level systems}\label{sec:3qubit}
\vspace{-3mm}
In this section, we now consider three effective TLSs with Rabi frequencies of $\Omega$, $\sqrt{2}\,\Omega$, and $\sqrt{3}\,\Omega$ (i.e., $n_1=1$, $n_2=2$ and $n_3=3$), whose study applies to the case of three-qubit gates and to that of systems with three sets of clusters with respectively $1$, $2$, and $3$ atoms (extensions to other Rabi frequency values are straightforward). The state variables are $\theta_k$, $\phi_k$, and $\phi_{g,k}$, with $k \in \{1,2,3\}$.

Here, we focus on two main processes, the controlled-controlled-Z (CCZ) gate and the simultaneous excitation process, which are of interest in the gate setting and in the cluster setting, respectively. 

First, we consider the CCZ gate, which is a three-qubit generalization of the CZ gate. In the computational subspace, the action of this gate reads: ${{\rm CCZ}=\mathbb{1}-2\ket{1,1,1}\bra{1,1,1}}$ (where $\mathbb{1}$ is the identity). We however use the following alternative definition, which is equivalent up to exchanging 0 and 1: ${{\rm CCZ}=2\ket{0,0,0}\bra{0,0,0}-\mathbb{1}}$. With this definition, in the three-TLS formalism, the CCZ gate corresponds (up to a single-qubit phase) to the following process:
\begin{align}
    \begin{split}
        \ket{g,1}&\rightarrow -e^{i\alpha}\ket{g,1},\\
        \ket{g,2}&\rightarrow -e^{i(2\alpha)}\ket{g,2},\\
        \ket{g,3}&\rightarrow -e^{i(3\alpha)}\ket{g,3},
    \end{split}
\end{align}
with $\alpha$ free \footnote{The canonical CCZ gate would correspond to a different process on the three TLSs, namely $\ket{g,1}\to e^{i\alpha}\ket{g,1}$, $\ket{g,2}\to e^{i(2\alpha)}\ket{g,2}$, $\ket{g,3}\to-e^{i(3\alpha)}\ket{g,3}$ (with $\alpha$ free), which has a longer optimal time.}.

Second, we consider the simultaneous excitation process, analogous to that of the two-TLS case: it consists in bringing all three effective TLSs from their ground to their excited states. The corresponding evolution is thus:
\begin{align}
    \begin{split}
        \ket{g,1}&\rightarrow e^{i\alpha}\ket{e,1},\\
        \ket{g,2}&\rightarrow e^{i\beta}\ket{e,2},\\
        \ket{g,3}&\rightarrow e^{i\gamma}\ket{e,3},
    \end{split}
\end{align}
with $(\alpha,\beta,\gamma)$ free.

As in the previous section, we thus distinguish two cases, depending on whether the ground state phases $\phi_{g,k}$ matter for the target process, or not. Similarly to the $N=2$ case, a change of variables allows us to simplify the equation by removing the irrelevant degrees of freedom. Introducing $\phi_{21}=\phi_2-\phi_1$ and $\phi_{31}=\phi_3-\phi_1$, ${\Phi_2=\phi_{g,2}-2\phi_{g,1}}$, and $\Phi_3=\phi_{g,3}-3\phi_{g,1}$, Eq.~\eqref{eq:HJBgeneral} (with $N=3$) can be reduced to:
\begin{widetext}
\begin{multline} \label{eq:HJB3qgates}
1=\Omega\left\lvert i\frac{\partial J}{\partial \theta_1}+\frac{1}{{\rm tan}(\theta_1)}\left(\frac{\partial J}{\partial \phi_{21}}+\frac{\partial J}{\partial \phi_{31}}\right)+\sqrt{2}ie^{i\phi_{21}}\frac{\partial J}{\partial \theta_2}-\frac{\sqrt{2}e^{i\phi_{21}}}{{\rm tan}(\theta_2)}\frac{\partial J}{\partial \phi_{21}}+\sqrt{3}ie^{i\phi_{31}}\frac{\partial J}{\partial \theta_3}-\frac{\sqrt{3}e^{i\phi_{31}}}{{\rm tan}(\theta_3)}\frac{\partial J}{\partial \phi_{31}}\right. \\
\left.+\frac{1}{2{\rm tan}(\theta_1/2)}\left(2\frac{\partial J}{\partial \Phi_2}+3\frac{\partial J}{\partial \Phi_3}\right)-\frac{\sqrt{2}e^{i\phi_{21}}}{2{\rm tan}(\theta_2/2)}\frac{\partial J}{\partial \Phi_2}-\frac{\sqrt{3}e^{i\phi_{31}}}{{2\rm tan}(\theta_3/2)}\frac{\partial J}{\partial \Phi_3}\right\rvert.
\end{multline}
\end{widetext}
When ground state phases are irrelevant, this equation simplifies by removing the last three terms and the associated variables, $\Phi_2$ and $\Phi_3$.
The state vector corresponding to the initial state, i.e., to all three TLSs in their ground state (see Sec.~\ref{sec:HJBeqs2tls}), is $\vec{x}_0\equiv(\theta_{1,0},\theta_{2,0},\theta_{3,0},\phi_{21,0},\phi_{31,0},\Phi_{2,0},\Phi_{3,0})=(\pi,\pi,\pi,0,0,0,0)$ (removing the last two variables if the ground state phases are irrelevant).
\begin{figure}
    \centering
    \includegraphics[width=\linewidth]{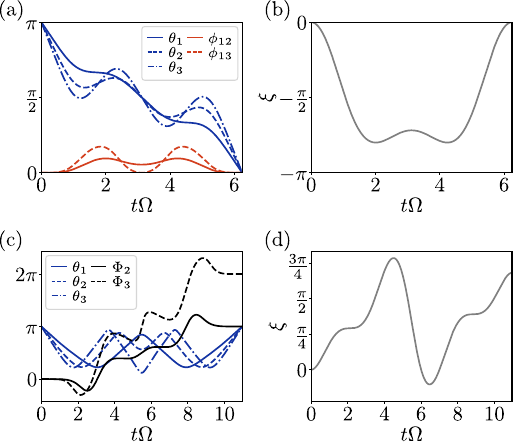}
    \caption{Optimal control results for three effective TLSs. (a) Optimal trajectory corresponding to the simultaneous excitation process: the three TLSs evolve from their ground state (${\theta_1=\theta_2=\theta_3=\pi}$) to their excited state (${\theta_1=\theta_2=\theta_3=0}$). (b) Optimal control profile associated to this trajectory, reconstructed via Eq.~\eqref{eq:optictrl}. (c) Optimal trajectory of the three-qubit CCZ entangling gate (${\rm CCZ}=2\ket{0,0,0}\bra{0,0,0}-\mathbb{1}$). The relative longitudes $\phi_{12}$ and $\phi_{13}$ are omitted from the plot. The black full and dashed lines represent the ground state phases (up to single-qubit rotations) $\Phi_2$ and $\Phi_3$, whose final values are respectively $\pi$ and $2\pi$, realizing the CCZ gate. (d) Optimal control profile associated to this trajectory.}
    \label{fig:3atoms}
\end{figure}

We find optimal times and controls for three-TLS processes by integrating the characteristic equations of Eq.~\eqref{eq:HJB3qgates}, resorting to the same short-time analysis presented in the previous section (and discussed in Appendix~\ref{App:shorttime}). Now, to find the optimal trajectory associated to a point $\vec{x}_t$, we must perform a search in a 4-dimensional (without ground state phases) or 6-dimensional (with ground state phases) parameter space, which is computationally harder than in the two-TLS case. Although ensuring global optimality is a more difficult task in this case, we are still able to find (at least locally) optimal trajectories for the simultaneous excitation process, and for the CCZ entangling gate. These results are shown in Fig.~\ref{fig:3atoms}.

For the simultaneous excitation process (i.e., ${\vec{x}_t=(0,0,0,0,0)}$), we find an optimal time of $T=6.25/\Omega$ (see Fig.~\ref{fig:3atoms}(a,b)). In the CCZ gate case (i.e., ${\vec{x}_t=(\pi,\pi,\pi,0,0,\pi,0)}$), we find $T=10.96/\Omega$ (see Fig.~\ref{fig:3atoms}(c,d)). We note that the optimal-time pulse derived for the CCZ gate, although slightly different, shows excellent agreement with the ansatz used in Ref.~\cite{everedHighFidelityParallel}.

\section{Conclusion}
\vspace{-3mm}
We applied the Bellman principle of optimality to address time-optimal control problems in Rydberg-atom systems. By leveraging the Bloch sphere representation of effective qubits valid in the blockade limit, we derived a minimum-dimensional expression of the associated Hamilton-Jacobi-Bellman equation, whose generalized solution corresponds to the global optimum. 
This approach formally rephrases time-optimal control problems for $N$-qubit blockade processes into the solution of a nonlinear, $M$-dimensional PDE, with $M=3N-2$ or $M=2N-1$ (depending on the target process). We used a generalization of the method of characteristics to find the solution: for a given target process, this corresponds to searching for the optimal trajectory in a parameter space of dimension $M-1$.  
We demonstrated this approach by optimizing several multiqubit processes relevant to diverse quantum computing settings, including controlled-phase gates and cluster dynamics. Interestingly, many of the time-optimal controls are interpretable, in the sense that they motivate simple ans{\"a}tze for control pulses that can be analytically solved and achieve similar durations. 

From a computational standpoint, the method developed in this work requires different resources compared to strategies based on direct or variational optimization of the pulse shape. 
It trades the (in principle infinite-dimensional) pulse optimization for an optimal characteristics search over a $2(N-1)$- or $3(N-1)$-dimensional parameter space. The linear scaling of the dimension of the search space with the number of qubits suggests that the method developed in this work can be similarly applied to more general $N$-qubit processes (e.g., optimal $\rm{C^{(n)}Z}$ gates), and thus adds to the toolbox for designing optimal dynamics for Rydberg systems.

\section*{Acknowledgements}
\vspace{-3mm}
We thank Giuliano Giudici for helpful discussions. 
This work is supported by the ERC Starting Grant QARA (Grant No.~101041435), and by the European Union's Horizon Europe research and innovation program under Grant Agreement No.~101113690 (PASQuanS2.1).

\appendix

\section{Reference frame and laser detuning}\label{App:phasevsdetuning}
\vspace{-3mm}

Here, we comment on the choice of reference frame corresponding to the Hamiltonians in Eqs.~(\ref{eq:hamiltonian},\ref{eq:hamiltonian_gr}). In doing so, we highlight the physical equivalence between considering the laser phase (as done in the main text) or the laser detuning as the control parameter.

In the lab frame, the dynamics of an atom, driven by a classical field tuned near the atomic transition $\omega_{eg}$ between two levels $\ket{g}$ and $\ket{e}$, is described in the rotating wave approximation by the following time-dependent Hamiltonian:
\begin{multline}
\label{eq:labframe}
H(t)=\Omega(t)\left(e^{i(\omega_{ge}t+\xi(t))}\ket{g}\bra{e} +e^{-i(\omega_{ge}t+\xi(t))}\ket{e}\bra{g}\right)\\+\omega_{ge}\ket{e}\bra{e},
\end{multline}
where $\Omega(t)$ and $\xi(t)$ indicate the Rabi frequency and the phase of the driving field, respectively.

This Hamiltonian is then simplified by moving to an appropriately rotating reference frame. For a change of frame given by a unitary operator $U(t)$ (i.e., $\ket{\psi'}=U(t)\ket{\psi}$), the Hamiltonian is transformed as ${H'(t)=U(t)H(t)U^{\dagger}(t)+i\dot{U}(t)U^{\dagger}(t)}$. In the case of Eq.~\eqref{eq:labframe}, two choices of rotating frame are possible. On the one hand, by considering the unitary transformation ${U_a(t)=e^{i\omega_{ge}t}\ket{e}\bra{e}+\ket{g}\bra{g}}$, the lab frame Hamiltonian transforms to:
\begin{align}
\label{eq:Hamiltonian_phase}
    H_a(t)=\frac{\Omega(t)}{2}\left(e^{i\xi(t)}\ket{g}\bra{e}+{\rm h.c.}\right),
\end{align}
which corresponds to the choice made in the main text.
On the other hand, by considering ${U_b(t)=e^{i(\omega_{ge}t+\xi(t)-\xi(0))}\ket{e}\bra{e}+\ket{g}\bra{g}}$, 
the lab frame Hamiltonian transforms to:
\begin{align}
\label{eq:Hamiltonian_detuning}
    H_b(t)=\frac{\Omega(t)}{2}\left(e^{i\xi(0)}\ket{g}\bra{e}+{\rm h.c.}\right)-\frac{\dot{\xi}(t)}{2}\left(\sigma_z+\mathbb{1}\right),
\end{align}
with $\sigma_z=\ket{e}\bra{e}-\ket{g}\bra{g}$. 
Identifying the time-derivative of the laser phase with the laser detuning, i.e., ${\dot{\xi}(t)\equiv\Delta(t)}$, this is the Hamiltonian used to describe a two-level atom driven by a detuned laser, in the frame rotating with the laser frequency.

These two choices of frame are related by the unitary transformation ${U(t)=e^{i(\xi(t)-\xi(0))}\ket{e}\bra{e}+\ket{g}\bra{g}}$, i.e., ${H_b(t)=U(t)H_a(t)U^{\dagger}(t)+i\dot{U}(t)U^{\dagger}(t)}$. Switching between the two pictures, thus, simply implies a dynamical redefinition of the excited state by a phase factor $e^{i(\xi(t)-\xi(0))}$, which corresponds to moving to a reference frame rotating around the $\hat{z}$ axis with velocity $\dot{\xi}(t)$. We note that, when switching between the two frames, the expression of the quantum states, and thus the state trajectories on the Bloch spheres, are also modified accordingly.

\section{Derivation of the HJB equation}\label{App:HJB}
\vspace{-3mm}
Here, we detail the steps to rewrite the stationary HJB equation from its form in Eq.~\eqref{eq:HJBscPSI} to that in Eq.~\eqref{eq:HJBgeneral}. 
From  Eq.~\eqref{eq:Bloch} we obtain:
\begin{align}
    \theta_k&=2{\rm arctan}\left(\sqrt{\frac{\psi_{g,k}\overline{\psi}_{g,k}}{\psi_{e,k}\overline{\psi}_{e,k}}}\right),\\
    \phi_{g,k}&=-\frac{i}{2}\log\left(\frac{\psi_{g,k}}{\overline{\psi}_{g,k}}\right),\\
    \phi_{k}&=-\frac{i}{2}\left[\log\left(\frac{\psi_{g,k}}{\overline{\psi}_{g,k}}\right)-\log\left(\frac{\psi_{e,k}}{\overline{\psi}_{e,k}}\right)\right].
\end{align}
We note that a fourth variable is in principle required for invertibility (e.g., $r_k=\sqrt{\psi_{g,k}\overline{\psi}_{g,k}+\psi_{e,k}\overline{\psi}_{e,k}}$), which is an integral of motion ($\dot{r}_k = 0$); as such, it does not enter the HJB equation, and thus we do not consider it in the following. 
From the equations above, applying the chain rule, we obtain the following expressions for the partial derivatives:
\begin{align}
        \begin{split}\label{eq:appendixB_1}
        \frac{\partial J^*}{\partial \psi_{g,k}}&=e^{-i\phi_{g,k}}\left({\rm cos}(\theta_k/2)\frac{\partial J^*}{\partial \theta_k}\right.\\
        &\qquad\qquad\left.-\frac{i}{2{\rm sin}(\theta_k/2)}\left[\frac{\partial J^*}{\partial \phi_k}+\frac{\partial J^*}{\partial \phi_{g,k}}\right]\right),
        \end{split}
        \\
        \begin{split}\label{eq:appendixB_2}
        \frac{\partial J^*}{\partial \psi_{e,k}}&=e^{i(\phi_k-\phi_{g,k})}\left(-{\rm sin}(\theta_k/2)\frac{\partial J^*}{\partial \theta_k}\right.\\
        &\qquad\qquad\left.+\frac{i}{2{\rm cos}(\theta_k/2)}\frac{\partial J^*}{\partial \phi_k}\right),
        \end{split}
\end{align}
and $\partial J^*/\overline{\psi}_{g,k}$ and $\partial J^*/\overline{\psi}_{e,k}$ are equal to the complex conjugate of the right-hand side of Eqs.~(\ref{eq:appendixB_1},\ref{eq:appendixB_2}). Substituting these expressions into Eq.~\eqref{eq:HJBscPSI}, and rearranging the terms (using the relation ${\tan(x/2)-1/\tan(x/2)=-2/\tan(x)}$), we obtain $1=\Omega \max_{\xi}\left[{\rm M}(\xi)\right]$, where 
\begin{align}
\begin{split}
{\rm M}(\xi)&=\sum_{k}\sqrt{n_k}\left({\rm sin}(\phi_{k}-\xi)\frac{\partial J^*}{\partial \theta_k}\right.\\
&\left.+{\rm cos}(\phi_{k}-\xi)\left[ \frac{1}{{\rm tan}(\theta_k)}\frac{\partial J^*}{\partial \phi_{k}}+\frac{1}{2{\rm tan}(\theta_k/2)}\frac{\partial J^*}{\partial \phi_{g,k}}\right]\right).
\end{split}
\end{align}
Then, expanding sine and cosine in terms of complex exponential functions, we obtain:
\begin{align}
\begin{split}
    {\rm M}(\xi)&={\rm Re}\left[e^{-i\xi}\sum_{k=1}^{N}\sqrt{n_k}e^{i\phi_k}\left(-i\frac{\partial J^*}{\partial \theta_k}+ \frac{1}{{\rm tan}(\theta_k)}\frac{\partial J^*}{\partial \phi_{k}}\right.\right.\\
&\left.\left.\qquad\qquad+\frac{1}{2{\rm tan}(\theta_k/2)}\frac{\partial J^*}{\partial \phi_{g,k}}\right)\right]\equiv {\rm Re}\left[e^{-i\xi}{\rm A}\right],
\end{split}
\end{align}
which can be analytically maximized over $\xi$ by setting $\xi={\rm arg} \left[{\rm A}\right]$, leading to Eq.~\eqref{eq:optictrl}. It thus follows that $1=\Omega \max_{\xi}\left[{\rm M}(\xi)\right]=\Omega \left|{\rm A}\right|$, which is indeed the HJB equation in Eq.~\eqref{eq:HJBgeneral}.

\section{Details on the generalized method of characteristics}\label{App:characsmethod}
Here, we provide a more detailed description of the method outlined in Sec.~\ref{sec:characteristics}.

The characteristics of a first-order PDE of the type ${F(\vec{x},\vec{p},J^*)=0}$, where $\vec{p}\equiv\nabla_{\vec{x}}J^*$, are defined as the solutions of the following system of ordinary differential equations (ODEs): $\dot{\vec{x}}(s)=\nabla_{\vec{p}}F$, ${\dot{\vec{p}}(s)=-\nabla_{\vec{x}}F-(\partial F/\partial J^*)\vec{p}}$, and $\dot{J^*}(s)=\vec{p}\cdot\nabla_{\vec{p}}F$. From their knowledge, it is possible to rebuild the graph of the solution of the PDE as follows: given a characteristic ${\gamma:s\in[0,\infty)\to(\vec{x}_{\gamma}(s),\vec{p}_{\gamma}(s),J^*_{\gamma}(s))}$, it results that $J^*(\vec{x}_{\gamma}(s))=J^*_{\gamma}(s)$. When characteristics cross one another in the state space, however, this method fails.

As discussed in Sec.~\ref{sec:characteristics}, as a consequence of its nonlinear nature, the characteristics of the HJB equation in Eq.~\eqref{eq:HJBgeneral} do cross. The emergence of such crossing regions, known as shocks, reflects the absence of a smooth solution of the PDE, which indeed admits a solution only in the generalized viscosity sense. Such a generalized solution can be computed by means of a generalization of the method of characteristics, grounded in an entropy principle that states that new characteristics cannot be generated at the crossings~\cite{sethianOrderedUpwind}. At each point $\vec{x}$ in the state space, $J^*(\vec{x})$ is thus computed as the smallest of the values of $J^*$ among the ones obtained from the characteristics passing through the point. Formally:
\begin{equation}
    {J^*(\vec{x})=\min_{\gamma\in\Gamma_{\vec{x}}}\left\{J^*_{\gamma}(s_x), \text{ with } s_x \text{ s.t. } \vec{x}_{\gamma}(s_x)=\vec{x}\right\}},
\end{equation}
where $\Gamma_{\vec{x}}$ is defined as the set of all characteristics passing through $\vec{x}$.

The target of the time-optimal control problem defines the boundary condition to assign to the associated HJB equation. From this boundary condition, the initial conditions of the characteristics' ODE system are then derived. Taking $M$ to be the dimension of the state space, in the general case of determining the minimum time to reach a hypersurface $\Sigma$ of dimension $M-1$, the boundary condition to assign to the HJB equation is $J^*\rvert_{\Sigma}=0$. The characteristics of interest are thus the ones starting from $\Sigma$. The set of characteristics has dimension $M-1$ and it is indexed only by $\vec{x}_0\in\Sigma$. Indeed, the initial momentum $\vec{p}_0$ of a characteristic starting from $\vec{x}_0$ is automatically fixed by the orthogonality condition (i.e., $\nabla_{\vec{x}} J^*(\vec{x})$ is orthogonal to the level sets of $J^*(\vec{x})$, and so $\vec{p}_0$ is orthogonal to $\Sigma$) and by the constraint ${F(\vec{x}_0,\vec{p}_0,0)=0}$ imposed by the HJB equation at the start of the characteristic. In the case of targeting a single point $\vec{x}_0$, the boundary condition to assign to the HJB equation is $J^*(\vec{x}_0)=0$, and the characteristics of interest are the ones starting from this point. In this case, the set of characteristics is indexed by the initial values of the momentum variables, $\vec{p}_0$, and it has dimension $M-1$ (due to the constraint $F(\vec{x}_0,\vec{p}_0,0)=0$ which removes one degree of freedom).

The HJB equation in Eq.~\eqref{eq:HJBgeneral} appears in the following form: ${F(\vec{x},\vec{p})=\Omega\left\lvert c_1(\vec{x})p_1+\dots+c_M(\vec{x})p_M\right\rvert}-1=0$, where the $c_i(\vec{x})$ (with $i\in\{1,\dots,M\}$) are given functions of the state variables only. For this class of HJB equations, the characteristic system returns ${\dot{J}^*(s)=1}$ which, together with the initial condition ${J^*(0)=0}$, gives ${J^*(s)=s}$. In addition, since $\dot{\vec{x}}=\nabla_{\vec{p}}F=-\vec{a}(\vec{x},\vec{u}^*)$, the state vector evolves according to the physical dynamics (given by $\vec{a}$) determined by the optimal control $\vec{u}^*$. 
Thus, the (state space projection of the) characteristics of Eq.~\eqref{eq:HJBgeneral} correspond to the physical optimal trajectories of the Rydberg system, and the parameter $s$ is simply the time variable along them. Moreover, before the characteristics cross, the level sets of $J^*$ can be computed by fixing $s$ in the characteristics: the set where $J^*(\vec{x})=A$ is given by $\left\{\vec{x}_{\gamma}(A),\:\gamma\in\Gamma\right\}$, with $\Gamma$ being the set of all characteristics.

Given an optimal characteristic $\gamma(s)$, the associated control $\xi^*_{\gamma}(s)$, generating the optimal trajectory $\vec{x}_{\gamma}(s)$, can be directly computed from Eq.~\eqref{eq:optictrl}. By looking at $\xi^*$ as a function of both coordinates and momenta  ${\xi^*(\vec{x},\vec{p})={\rm arg}\left\{c_1(\vec{x})p_1+\dots+c_M(\vec{x})p_M\right\}-\pi}$, it indeed results $\xi^*_{\gamma}(s)=\xi^*\left({\vec{x}_{\gamma}(s),\vec{p}_{\gamma}}(s)\right)$. From this, it is also possible to analytically derive necessary conditions that the control should satisfy for optimality. It is for instance possible to compute the initial slope of the optimal laser phase (i.e., the optimal initial detuning), which we use in Sec.~\ref{Sec:results}. Indeed, $\dot{\xi}^*_{\gamma}(s)$ can be computed by evaluating the Poisson bracket between $\xi^*(\vec{x},\vec{p})$ and $F(\vec{x},\vec{p})$ on $\gamma(s)$ (i.e., at $\vec{x}=\vec{x}_{\gamma}(s)$ and $\vec{p}=\vec{p}_{\gamma}(s)$):
\begin{equation}\label{eq:PoissonAppendix}
    \dot{\xi}^*_{\gamma}(s)=\left.\left\{\xi^*,F\right\}\right|_{\gamma(s)}\equiv\left.\nabla_{\vec{x}}\xi^*\cdot\nabla_{\vec{p}}F-\nabla_{\vec{p}}\xi^*\cdot\nabla_{\vec{x}}F\right|_{\gamma(s)}.
\end{equation}
The initial slope of the optimal laser phase is thus obtained by taking the $s\to0$ limit in Eq.~\eqref{eq:PoissonAppendix}. Similarly, it is possible to compute the initial higher-order derivatives of the laser phase, e.g., ${\ddot{\xi}^*_{\gamma}(0)=\lim_{s\to0}\left.\left\{\left\{\xi^*,F\right\},F\right\}\right|_{\gamma(s)}}$. 

\section{Analytic expressions of the characteristics in the single-qubit case}\label{App:analytic}
\vspace{-3mm}
Here, we detail the analytic solution of the characteristic equations of the single-qubit time-optimal control problem (Eq.~\eqref{eq:characs1qubit}):
\begin{align}
        \frac{d\theta}{ds}&=\Omega^2 p_{\theta},\label{eq:theq}\\
        \frac{d\phi}{ds}&=\Omega^2 \frac{p_{\phi}}{{\rm tan}^2(\theta)},\label{eq:phieq}\\
        \frac{dp_{\theta}}{ds}&=\Omega^2 \frac{p_{\phi}^2}{{\rm tan}(\theta){\rm sin}^2(\theta)},\label{eq:p1eq}\\
        \frac{dp_{\phi}}{ds}&=0,\label{eq:p2eq}
\end{align}
with ${\left(\theta(0),\phi(0),p_{\theta}(0),p_{\phi}(0)\right)=\left(\theta_0,\phi_0,p_{\theta,0},p_{\phi_0}\right)}$ as initial conditions. The HJB equation gives the following constraint for the initial momenta $p_{\theta,0}^2+p_{\phi,0}^2{\rm cot}^2(\theta_0)=1/\Omega^2$, and from Eq.~\eqref{eq:p2eq} we get that $p_{\phi}$ remains constant on the characteristic. Plugging Eq.~\eqref{eq:p1eq} in the derivative of Eq.~\eqref{eq:theq}, we obtain:
\begin{align}
    \frac{d^2\theta}{ds^2}&=\Omega^4 \frac{p_{\phi}^2}{{\rm tan}(\theta){\rm sin}^2(\theta)},
\end{align}
from which we perform the following steps:
\begin{align}
    \frac{d\theta}{ds}\frac{d^2\theta}{ds^2}&=-\Omega^4 \frac{d\theta}{ds}p_{\phi}^2{\rm cot}(\theta)\frac{d}{d\theta}\left({\rm cot}(\theta)\right),\nonumber\\
    \frac{1}{2}\frac{d}{ds}\left[\left(\frac{d\theta}{ds}\right)^2\right]&= -\frac{\Omega^4 p_{\phi}^2}{2}\frac{d}{d\theta}\left({\rm cot}^2(\theta)\right)\frac{d\theta}{ds},\nonumber\\ \label{eq:AppendixC_1}
    \left(\frac{d\theta}{ds}\right)^2&= -\Omega^4 p_{\phi}^2{\rm cot}^2(\theta)+{\rm cst}.
\end{align}
From Eq.~\eqref{eq:theq} at $s=0$, using the constraint on the initial momenta, we get $(d\theta/ds)^2(0)=
\Omega^2-\Omega^4 p_{\phi}^2{\rm cot}^2(\theta_0)$;
thus, the constant in Eq.~\eqref{eq:AppendixC_1} is equal to $\Omega^2$. It follows that:
\begin{align}
    \frac{d\theta}{ds}&= \sigma\Omega\sqrt{1-\Omega^2 p_{\phi}^2{\rm cot}^2(\theta)},\nonumber\\
    {\rm sin}(\theta)\frac{d\theta}{ds}&=\sigma\Omega\sqrt{{\rm sin}^2(\theta)-\Omega^2 p_{\phi}^2{\rm cos}^2(\theta)},\nonumber\\ 
    -\frac{d({\rm cos}(\theta))}{ds}&=\sigma\Omega\sqrt{1-(1+\Omega^2 p_{\phi}^2){\rm cos}^2(\theta)},
\end{align}
with $\sigma=\pm 1$. Given the change of variable ${u=A{\rm cos}(\theta)}$, with ${A=\sqrt{1+\Omega^2 p_{\phi}^2}}$, it follows that:
\begin{equation}\label{eq:AppendixC_2}
    \sigma\Omega A=-\frac{1}{\sqrt{1-u^2}}\frac{du}{ds}=\frac{d({\rm arccos}(u))}{ds},
\end{equation}    
and then: 
\begin{equation}
    {\rm arccos}\left(A{\rm cos}(\theta)\right)=\pm\Omega s A+{\rm arccos}(A{\rm cos}(\theta_0)),
\end{equation}
having employed the initial condition $\theta(0)=\theta_0$. From this, we eventually get:
\begin{align}\label{eq:thetaofs}
    \theta(s)={\rm arccos}\left( \frac{{\rm cos}\left(\sigma\Omega s A+{\rm arccos}(A{\rm cos}(\theta_0))\right)}{A}\right).
\end{align}
We now exploit this expression to compute $\phi(s)$. Posing ${B(s)\equiv\sigma\Omega s A+{\rm arccos}(A{\rm cos}(\theta_0))}$,  Eq.~\eqref{eq:phieq} reads:
\begin{align}
    \frac{d\phi}{ds}&=\frac{\Omega^2 p_{\phi}{\rm cos}^2\left(B(s)\right)}{A^2-{\rm cos}^2\left(B(s)\right)}\nonumber\\
    &=\Omega^2 p_{\phi}\left(-1+\frac{A^2 {\rm sec}^2(B(s))}{A^2(1+{\rm tan}^2(B(s)))-1}\right)\nonumber\\
    &=-\Omega^2 p_{\phi}+\sigma\frac{d}{ds}\left( {\rm arctan}\left(\frac{A}{\Omega p_{\phi}}{\rm tan}(B(s))\right)\right),
\end{align}
from which it follows:
\begin{align}\label{eq:phiofs}
    \begin{split}
    \phi(s)&=-\Omega^2 p_{\phi} s +\sigma{\rm arctan}\left(\frac{A}{\Omega p_{\phi}}{\rm tan}(B(s))\right)\\
    &+\phi_0-\sigma{\rm arctan}\left(\frac{\sqrt{1-A^2{\rm cos}^2(\theta_0)}}{\Omega p_{\phi}{\rm cos}(\theta_0)}\right),
    \end{split}
\end{align}
having imposed the initial condition $\phi(0)=\phi_0$.

Eqs.~(\ref{eq:thetaofs},\ref{eq:phiofs}) give the general expression of $\theta(s)$ and $\phi(s)$, for an arbitrary target point, and the result of the main text (Eq.~\eqref{eq:characteristicsEquator}) is retrieved by setting ${\left(\theta_0,\phi_0\right)=\left(\pi/2,0\right)}$.

The optimal control can then be computed via Eq.~\eqref{eq:optictrl}, giving $\xi^*(s)=-\Omega^2p_{\phi}s+\phi_0\pm {\rm arccos}(p_{\phi}{\rm cot}(\theta_0))$. Alternatively, this can also be obtained from its derivative $\dot{\xi}(s)$, which is equal to (see Eq.~\eqref{eq:PoissonAppendix}):
\begin{align}\label{eq:dxistar}
    \begin{split}
    \frac{d\xi^*}{ds}&=\frac{d\theta}{ds}\partial_{\theta}\xi^*+\frac{d\phi}{ds}\partial_{\phi}\xi^*+\frac{d p_{\theta}}{ds}\partial_{p_{\theta}}\xi^*=-p_{\phi}\Omega^2,
    \end{split}
\end{align}
implying that $\xi^*(s)=-\Omega^2p_{\phi}s+{\rm cst}$. The constant is determined from the initial conditions via Eq.~\eqref{eq:optictrl}.

\section{Derivation of the two-TLS HJB equations}\label{App:HJB2tls}
\vspace{-3mm}
Here, we detail how Eqs.~\eqref{eq:HJBampli} and~\eqref{eq:HJBgates} are derived from Eq.~\eqref{eq:HJBgeneral} in the case of two TLSs with ${n_1=1}$ and ${n_2=2}$. Performing the change of variables ${(\phi,\tilde{\phi})=(\phi_2-\phi_1,\phi_1+\phi_2)}$, Eq.~\eqref{eq:HJBgeneral} reads:
\begin{align}
\label{eq:AppendixD_1}
1&=\Omega\left\lvert i\frac{\partial J^*}{\partial \theta_1}+\sqrt{2}ie^{i\phi}\frac{\partial J^*}{\partial \theta_2}+\left(\frac{1}{{\rm tan}(\theta_1)}-\frac{\sqrt{2}e^{i\phi}}{{\rm tan}(\theta_2)}\right)\frac{\partial J^*}{\partial \phi} \right.\notag\\&-\left(\frac{1}{{\rm tan}(\theta_1)}+\frac{\sqrt{2}e^{i\phi}}{{\rm tan}(\theta_2)}\right)\frac{\partial J^*}{\partial \tilde{\phi}}- \frac{1}{2{\rm tan}(\theta_1/2)}\frac{\partial J^*}{\partial \phi_{g,1}} \notag\\
&\quad\left. -\frac{e^{i\phi}}{\sqrt{2}{\rm tan}(\theta_2/2)}\frac{\partial J^*}{\partial \phi_{g,2}}\right\rvert.
\end{align}
Throughout this work, we consider targets that are independent of the value of $\tilde{\phi}$ (which corresponds to the average Bloch sphere longitude of the two TLSs). As $\tilde{\phi}$ also does not explicitly appear in the HJB equation in Eq.~\eqref{eq:AppendixD_1}, it follows that $\partial J^*/\partial\tilde{\phi}=0$ everywhere, and this variable can be eliminated from the equation altogether.

Similarly, $\phi_{g,1}$ and $\phi_{g,2}$ do not explicitly appear in the equation and thus we can distinguish two cases depending on whether the optimal control problem involves the ground-state phases or not. In the latter case, since the boundary condition does not depend on $\phi_{g,1}$ and $\phi_{g,2}$, we have that $\partial J^*/\partial \phi_{g,1}=\partial J^*/\partial \phi_{g,2}=0$ everywhere, and Eq.~\eqref{eq:AppendixD_1} reduces to Eq.~\eqref{eq:HJBampli}. In the former case, we are usually interested in gate operations ``up to a single-qubit rotation": the quantity of interest is the gate phase $\Phi\equiv\phi_{g,2}-2\phi_{g,1}$, and $\phi_{g,1}$ is interpreted as a single-qubit rotation. Performing the change of variables $(\Phi,\tilde{\Phi})=(\phi_{g,2}-2\phi_{g,1},\phi_{g,2}+2\phi_{g,1})$, we can then eliminate $\tilde{\Phi}$ (since it appears neither in the equation nor in the boundary condition) and eventually derive Eq.~\eqref{eq:HJBgates}.

\section{Expressions of the characteristic equations for two TLS}\label{App:characs}
\vspace{-3mm}
Here, we give the characteristic equations associated to Eqs.~\eqref{eq:HJBampli} and~\eqref{eq:HJBgates}. The characteristic equations associated to Eq.~\eqref{eq:HJBgates} are:
\begin{widetext}
\begingroup
\allowdisplaybreaks
\begin{align}
\label{eq:AppendixE_1}
    \frac{d\theta_1}{ds}&=\Omega^2 \left[p_{\theta_1}+\sqrt{2}{\rm cos}(\phi)p_{\theta_2}-\sqrt{2}\frac{{\rm sin}(\phi)}{{\rm tan}(\theta_2)}p_{\phi}-\frac{{\rm sin}(\phi)}{\sqrt{2}{\rm tan}(\theta_2/2)}p_{\Phi}\right],\\
    \frac{d\theta_2}{ds}&=\Omega^2 \left[\sqrt{2}{\rm cos}(\phi)p_{\theta_1}+2p_{\theta_2}-\sqrt{2}\frac{{\rm sin}(\phi)}{{\rm tan}(\theta_1)}p_{\phi}-\sqrt{2}\frac{{\rm sin}(\phi)}{{\rm tan}(\theta_1/2)}p_{\Phi}\right],\\
    \label{eq:AppendixE_3}\begin{split}\frac{d\phi}{ds}&=\Omega^2 \left[\left(\frac{1}{{\rm tan}^2(\theta_1)}+\frac{2}{{\rm tan}^2(\theta_2)}-\frac{2\sqrt{2}{\rm cos}(\phi)}{{\rm tan}(\theta_1){\rm tan}(\theta_2)}\right)p_{\phi} -\sqrt{2}{\rm sin}(\phi)\left(\frac{p_{\theta_2}}{{\rm tan}(\theta_1)}+\frac{p_{\theta_1}}{{\rm tan}(\theta_2)}\right)\right.\\
    &\quad\left.+\lr{\frac{1}{{\rm tan}(\theta_1/2){\rm tan}(\theta_1)} - \frac{{\rm cos}(\phi)}{\sqrt{2}{\rm tan}(\theta_1){\rm tan}(\theta_2/2)} -\frac{\sqrt{2}{\rm cos}(\phi)}{{\rm tan}(\theta_1/2){\rm tan}(\theta_2)} + \frac{1}{{\rm tan}(\theta_2/2){\rm tan}(\theta_2)} }p_{\Phi}\right],\end{split}\\
    \begin{split}\frac{d\Phi}{ds}&=\Omega^2 \left[\lr{\frac{1}{{\rm tan}^2(\theta_1/2)} +\frac{1}{2{\rm tan}^2(\theta_2/2)} - \frac{\sqrt{2}{\rm cos}(\phi)}{{\rm tan}(\theta_1/2){\rm tan}(\theta_2/2)}}p_{\Phi}-\frac{{\rm sin}(\phi)}{\sqrt{2}{\rm tan}(\theta_2/2)}p_{\theta_1} -\frac{\sqrt{2}{\rm sin}(\phi)}{{\rm tan}(\theta_1/2)}p_{\theta_2} \right. \\
    &\quad \left. + \lr{\frac{1}{{\rm tan}(\theta_1/2){\rm tan}(\theta_1)} - \frac{{\rm cos}(\phi)}{\sqrt{2}{\rm tan}(\theta_1){\rm tan}(\theta_2/2)} -\frac{\sqrt{2}{\rm cos}(\phi)}{{\rm tan}(\theta_1/2){\rm tan}(\theta_2)} + \frac{1}{{\rm tan}(\theta_2/2){\rm tan}(\theta_2)} }p_{\phi}\right],\end{split}\\
    \label{eq:AppendixE_5}\begin{split}\frac{dp_{\theta_1}}{ds}&=\Omega^2 \left[\left(\frac{1}{{\rm tan}(\theta_1){\rm sin}^2(\theta_1)}-\frac{\sqrt{2}{\rm cos(\phi)}}{{\rm tan}(\theta_2){\rm sin}^2(\theta_1)}\right)p_{\phi}^2 -\sqrt{2}\frac{{\rm sin}(\phi)}{{\rm sin}^2(\theta_1)}p_{\theta_2}p_{\phi}\right. \\
    &\left. \quad +\left(\frac{1}{2{\rm tan}(\theta_1/2){\rm sin}^2(\theta_1/2)}-\frac{{\rm cos}(\phi)}{2\sqrt{2}{\rm tan}(\theta_2/2){\rm sin}^2(\theta_1/2)}\right)p_{\Phi}^2 -\frac{{\rm sin}(\phi)}{\sqrt{2}{\rm sin}^2(\theta_1/2)}p_{\theta_2}p_{\Phi}\right.\\
    &\left. \quad +\left(\frac{1}{2{\rm sin}^2(\theta_1/2){\rm tan}(\theta_1)}+\frac{1}{{\rm sin}^2(\theta_1){\rm tan}(\theta_1/2)}-\frac{{\rm cos}(\phi)}{\sqrt{2}{\rm sin}^2(\theta_1){\rm tan}(\theta_2/2)}-\frac{{\rm cos}(\phi)}{\sqrt{2}{\rm sin}^2(\theta_1/2){\rm tan}(\theta_2)}\right)p_{\phi}p_{\Phi}\right],\end{split}\\
    \begin{split}\frac{dp_{\theta_2}}{ds}&=\Omega^2 \left[\left(\frac{2}{{\rm tan}(\theta_2){\rm sin}^2(\theta_2)}-\frac{\sqrt{2}{\rm cos(\phi)}}{{\rm tan}(\theta_1){\rm sin}^2(\theta_2)}\right)p_{\phi}^2 -\sqrt{2}\frac{{\rm sin}(\phi)}{{\rm sin}^2(\theta_2)}p_{\theta_1}p_{\phi}\right. \\
    &\left. \quad +\left( \frac{1}{4{\rm tan}(\theta_2/2){\rm sin}^2(\theta_2/2)}-\frac{{\rm cos}(\phi)}{2\sqrt{2}{\rm tan}(\theta_1/2){\rm sin}^2(\theta_2/2)} \right)p_{\Phi}^2 -\frac{{\rm sin}(\phi)}{2\sqrt{2}{\rm sin}^2(\theta_2/2)}p_{\theta_1}p_{\Phi}\right.\\
    &\left. \quad +\left( -\frac{{\rm cos}(\phi)}{2\sqrt{2}{\rm sin}^2(\theta_2/2){\rm tan}(\theta_1)} - \frac{\sqrt{2}{\rm cos}(\phi)}{{\rm sin}^2(\theta_2){\rm tan}(\theta_1/2)} + \frac{1}{2{\rm sin}^2(\theta_2/2){\rm tan}(\theta_2)}+\frac{1}{{\rm sin}^2(\theta_2){\rm tan}(\theta_2/2)} \right)p_{\phi}p_{\Phi}\right],\end{split}\\
    \label{eq:AppendixE_7}\begin{split}\frac{dp_{\phi}}{ds}&=\sqrt{2}\Omega^2 \left[{\rm sin}(\phi)p_{\theta_1}p_{\theta_2}+{\rm cos}(\phi)\left(\frac{p_{\theta_1}}{{\rm tan}(\theta_2)}+\frac{p_{\theta_2}}{{\rm tan}(\theta_1)}\right)p_{\phi} -\frac{{\rm sin}(\phi)}{{\rm tan}(\theta_1){\rm tan}(\theta_2)}p_{\phi}^2 -\frac{{\rm sin}(\phi)}{\sqrt{2}{\rm tan}(\theta_1/2){\rm tan}(\theta_2/2)}p_{\Phi}^2 \right. \\
    &\left.\quad + \frac{{\rm cos}(\phi)}{\sqrt{2}{\rm tan}(\theta_2/2)}p_{\theta_1}p_{\Phi} +\frac{\sqrt{2}{\rm cos}(\phi)}{{\rm tan}(\theta_1/2)}p_{\theta_2}p_{\Phi} -\left( \frac{{\rm sin}(\phi)}{\sqrt{2}{\rm tan}(\theta_1){\rm tan}(\theta_2/2)} + \frac{\sqrt{2}{\rm sin}(\phi)}{{\rm tan}(\theta_1/2){\rm tan}(\theta_2)}\right)p_{\phi}p_{\Phi}\right],\end{split}\\
    \frac{dp_{\Phi}}{ds}&=0.\label{eq:AppendixE_8}
\end{align}
\end{widetext}\endgroup
The characteristic equations associated to Eq.~\eqref{eq:HJBampli} are obtained from those in Eqs.~(\ref{eq:AppendixE_1}-\ref{eq:AppendixE_3},\ref{eq:AppendixE_5}-\ref{eq:AppendixE_7}), by setting $p_{\Phi}$ to zero. 

\section{Detailed analysis of the short-time dynamics} \label{App:shorttime}
\vspace{-3mm}
Here, we detail the short-time analysis needed to compute the $s\sim0$ expressions of the state variables when all TLSs are driven from their ground state. The results of this analysis are used in Sec.~\ref{sec:HJB2tls}. 

Remembering that, for the HJB equations under analysis, the characteristics coincide with the physical trajectories, and the $s$ variable with the time $t$, we consider the dynamics of the effective TLSs at short time. 
The $k$-th effective TLS evolves under the time-dependent Hamiltonian $H_k(t)$ in Eq.~\eqref{eq:Hks}, with constant Rabi frequency $\Omega_k=\sqrt{n_k}\,\Omega$, and time-dependent phase $\xi(t)$. The time evolution of the initial state $\ket{\psi_k(0)}=\ket{g,k}$ can be computed via:
\begin{align}
\begin{split}
    \ket{\psi_k(t)}&=\sum_{n=0}^{\infty}(-i)^n\int_{0}^{t}d\tau_1\int_{0}^{\tau_1}d\tau_2\dots \int_{0}^{\tau_{n-1}}d\tau_n\\
    &\qquad\qquad\qquad\quad H_k(\tau_1)\dots H_k(\tau_n)\ket{g,k}.
\end{split}
\end{align}
By expanding the control phase at short time in power series as $\xi(t)=\xi_0+\xi_1 t +\xi_2 t^2 +\xi_3 t^3+O(t^4)$, the following power expansions for $\psi_{g,k}(t)\equiv\braket{g,k}{\psi_k(t)}$ and ${\psi_{e,k}(t)\equiv\braket{e,k}{\psi_k(t)}}$ result:
\begin{align}
\begin{split}
    &\psi_{g,k}(t)=1- \frac{\Omega_k^2 t^2}{8}-\frac{i\xi_1 \Omega_k^2 t^3}{24}\\
    &\quad+ \left(\frac{\Omega_k^4}{384}-\frac{i\xi_2\Omega_k^2}{24}+\frac{\xi_1^2 \Omega_k^2}{96}\right)t^4 +O\left(t^5\right),\\
    &\psi_{e,k}(t)=\Omega_k\left[-\frac{i t}{2} - \frac{\xi_1 t^2}{4} + \left(\frac{i\Omega_k^2}{48}-\frac{\xi_2}{6}+\frac{i\xi_1^2}{12}\right)t^3 \right.\\
    &\quad\left.+ \left(\frac{\xi_1^3}{48}+\frac{\xi_1\Omega_k^2}{96}+\frac{i\xi_1\xi_2}{8}-\frac{\xi_3}{8}\right)t^4\right]e^{-i\xi_0}+O\left(t^5\right).
\end{split}
\end{align}

Now, remembering that ${\theta_k(t)=\pi-2{\rm arcsin}\left(\left\lvert\psi_{e,k}(t)\right\rvert\right)}$, ${\phi_{g,k}(t)={\rm arg}[\psi_{g,k}(t)]}$, ${\phi_{e,k}(t)={\rm arg}[\psi_{e,k}(t)]}$, and ${\phi(t)=\phi_{g,k}(t)-\phi_{e,k}(t)}$, we obtain:
\begin{align}
    \theta_k(t)&=\pi-\Omega_k t+\frac{\xi_1^2 \Omega_k}{24} t^3+\frac{\xi_1\xi_2\Omega_k}{12}t^4+O(t^5),\\
    \phi_k(t)&=\frac{\pi}{2}+\xi_{0}+\frac{\xi_1 t}{2} +\frac{\xi_2 t^2}{3}+\frac{\left(6\xi_3-\xi_1\Omega_k^2\right)t^3}{24}+O(t^4).
\end{align}
In the case of two effective TLSs ($\Omega_1=\Omega$ and $\Omega_2=\sqrt{2}\,\Omega$), the state variables $\theta_1$, $\theta_2$ and $\phi=\phi_2-\phi_1$ admit the following short-time expressions:
\begin{align}
    \theta_1(t)&=\pi-\Omega t+\frac{\xi_1^2 \Omega}{24} t^3+O(t^4),\\
     \theta_2(t)&=\pi-\sqrt{2}\left(\Omega t-\frac{\xi_1^2 \Omega}{24} t^3\right)+O(t^4),\\
     \phi(t)&=-\frac{\xi_1 \Omega^2}{24}t^3+O(t^4).
\end{align}
These expressions justify the analysis conducted in Sec.~\ref{sec:HJB2tls}. Indeed, $\phi(t)$ visibly grows at least as a third power of $t$, ${\rm tan}(\theta_2)\sim\sqrt{2}{\rm tan}(\theta_1)\sim\sqrt{2}t$, and $(1/{\rm tan}(\theta_1))-(\sqrt{2}e^{i\phi}/{\rm tan}(\theta_2))=-\Omega t/3 + O\left(t^2\right)$.  

\section{Approximating optimal pulses with composite pulse sequences}\label{App:singlepulse}
\vspace{-3mm}
We now describe how the optimal pulses found in Sec.~\ref{Sec:results} for selective rotation processes (see Fig.~\ref{fig:thetas}) can be approximated by simple composite-pulse sequences, with a duration close to the optimal one.

The optimal times shown in Figs.~\ref{fig:thetas}(a)~and~\ref{fig:thetas}(b) display, depending on the target value, two qualitatively distinct behaviors. These two regimes are, in both plots, separated by a discontinuity in the derivative. We consider the case where the first TLS returns to its ground state while the second reaches some target $\theta_{2,\rm targ}$, and note that the same reasoning applies to the opposite case.

As discussed in Sec.~\ref{Sec:results}, to the right of the discontinuity, the optimal pulse can be approximated by a single pulse of constant detuning ($\xi(t)=\Delta t$). The relationship between the pulse duration $T$ and the target latitude $\theta_{2,\rm targ}$ can be derived analytically. Indeed, to ensure that the first effective TLS returns to its ground state, one requires the following (taking $\Omega=1$ for simplicity):
\begin{align}\label{eq:Tofdelta}
    T=\frac{2\pi}{\sqrt{1+\Delta^2}},
\end{align}
which can be inverted to:
\begin{align}\label{eq:deltaofT}
    \Delta=\pm\sqrt{\frac{4\pi^2}{T^2}-1}.
\end{align}
On the second Bloch sphere, this pulse induces a rotation (in a properly chosen rotating frame, see Appendix~\ref{App:phasevsdetuning}) of angle ${\theta_r=\sqrt{2+\Delta^2}T}$ around the axis defined by ${\hat{n}=(\sqrt{2}\hat{x}-\Delta\hat{z})/\sqrt{2+\Delta^2}}$ (where $\hat{x}$ and $\hat{z}$ are the unit vectors in the $x$ and $z$ directions, respectively). The ground state of the second TLS thus rotates to a final latitude, i.e., $\theta_{2,\rm targ}$, given by:
\begin{align}
    \theta_{2,\rm targ}&={\rm arccos}\left(\cos^2(\alpha)(1-\cos(\theta_r))-1\right),
\end{align}
where $\alpha={\rm arctan}(\Delta/\sqrt{2})$. Using Eqs.~(\ref{eq:Tofdelta},\ref{eq:deltaofT}), $\theta_{2,\rm targ}$ can be expressed either as a function of $\Delta$ or of $T$ and, for instance, it results:
\begin{align}
    \theta_{2,\rm targ}(T)&={\rm arccos}\left(\frac{T^2 \left[1- 2 \cos(\sqrt{4 \pi^2+T^2})\right]-4 \pi^2 }{
 4 \pi^2 + T^2}\right),
\end{align}
with $T\in\left[0,2\pi\right]$. This relationship between $\theta_{2,\rm targ}$ and $T$ is very close to the optimal one in the right side of the plot in Fig.~\ref{fig:thetas}(a). The point of discontinuity of the derivative corresponds to the resonant driving case (i.e., $\Delta=0$, $T=2\pi$).

To the left of the discontinuity, a single detuned pulse is no longer enough to reach the target. However, one can straightforwardly realize the process with a two-pulse sequence, consisting of two of the previously defined pulses (one resonant, one detuned) separated by a well-chosen phase jump. The first TLS returns to its ground state under both pulses and thus obviously under the sequence. The second TLS is sent by the resonant pulse to the latitude of the discontinuity,  $\theta_{2}=(3-2\sqrt{2})\pi$, and the rest of the way to $\theta_{2,\rm targ}$ is then completed by the second pulse. This two-pulse sequence is equivalent to a symmetric three-pulse sequence, whose shape resembles that of the optimal pulse. Indeed, we find that splitting the resonant pulse in two parts, placed around the detuned one, does not change the final latitude.

We note that the above line of reasoning applies to other processes as well. The shape of the optimal pulses can in general be used to guess efficient composite-pulse sequences that realize the same process with a near-optimal duration, allowing for an intuitive analytical description.

\bibliography{biblio.bib}
\end{document}